\documentclass[ba,preprint]{imsart}
\pubyear{2023}
\volume{TBA}
\issue{TBA}
\arxiv{2201.08502}
\firstpage{1}
\lastpage{1}

\usepackage{amsmath,amsthm,amssymb}
\usepackage{natbib}
\usepackage[colorlinks,citecolor=blue,urlcolor=blue,filecolor=blue,backref=page]{hyperref}
\usepackage{graphicx,graphics,float}
\usepackage{bm,bbm}
\usepackage{mathrsfs}

\usepackage[linesnumbered,plain,noend]{algorithm2e}
\usepackage[section]{placeins}
\usepackage[toc,page]{appendix}
\SetKwComment{Comment}{$\triangleright$\ }{}
\usepackage{epsfig}
\usepackage{subcaption, ragged2e}
\usepackage{tikz}
\usetikzlibrary{bayesnet}
\usetikzlibrary{arrows}
\theoremstyle{plain}
\newtheorem{prop}{Proposition}
\theoremstyle{plain}
\newtheorem{lemma}{Lemma}

\DeclareMathOperator{\diag}{diag}

\startlocaldefs

\endlocaldefs

\begin{document}


\begin{frontmatter}
\title{Curved factor analysis with the Ellipsoid-Gaussian distribution}
\runtitle{Ellipsoid-Gaussian}

\begin{aug}
\author{\fnms{Hanyu} \snm{Song}}
\thanksref{addr1,t1}\ead[label=e1]{hanyu.song@duke.edu}
\and
\author{\fnms{David} \snm{Dunson}\thanksref{addr1,t1}\ead[label=e2]{dunson@duke.edu}}
    
\runauthor{Song and Dunson}

\address[addr1]{ Department of Statistical Science,
    Duke University, Durham NC 27710, USA
     \printead{e1} 
    \printead*{e2}}
\thankstext{t1}{The authors gratefully acknowledge the support from the National Institute of Environmental Health Sciences of the United States National Institutes of Health.}

\end{aug}

\begin{abstract}
There is a need for new models for characterizing dependence in multivariate data. The multivariate Gaussian distribution is routinely used, but 
cannot characterize nonlinear relationships in the data.  Most non-linear extensions tend to be highly complex; for example, involving estimation of a non-linear regression model in latent variables. In this article, we propose a relatively simple class of Ellipsoid-Gaussian multivariate distributions, which are derived by using a Gaussian linear factor model involving latent variables having a von Mises-Fisher distribution on a unit hyper-sphere. We show that the Ellipsoid-Gaussian distribution can flexibly model curved relationships among variables with lower-dimensional structures. Taking a Bayesian approach, we propose a hybrid of gradient-based geodesic Monte Carlo and adaptive Metropolis for posterior sampling. We derive basic properties and illustrate the utility of the Ellipsoid-Gaussian distribution on a variety of simulated and real data applications. An accompanying R package is also available.
\end{abstract}

\end{frontmatter}


\section{Introduction}

The multivariate Gaussian distribution is routinely used, relying on a rich collection of methods for inference on the covariance structure. Factor analysis is particularly popular,
due to its combination of simplicity and flexibility.
The generic form of a Gaussian linear factor model is 
\begin{align}
    {x}_i = {c} + \Lambda {\eta}_i + \epsilon_i, \quad \epsilon_i \sim \text{N}_p(0, \Sigma),\quad (i = 1,\ldots, n), \label{equ:gaussfac}
\end{align}
where ${x}_i$ is $p$-dimensional, $\Lambda$ is a $p \times k$ factor loadings matrix, ${\eta}_i \sim \text{N}_k(0, I_k)$ are latent factors and $\epsilon_i$ is an idiosyncratic error with covariance $\Sigma = \diag(\sigma_1^2, \ldots, \sigma_p^2)$. Conditional on the factors, the elements of $x_i$ are independent; dependence is induced by marginalizing out the latent factors to obtain 
 ${x}_i \sim \text{N}_p({c}, \Omega)$ with $\Omega = \Lambda \Lambda^T + \Sigma$. 

In genomics, for instance, ${x}_i$ can be a massive vector of gene expression values; with genes within common pathways tending to co-express, it is natural to regard 
$\eta_i$ as pathway characterizing factors \citep{CCLNWM08}. In practice,
one typically chooses $k \ll p$. It is reasonable to suppose that the loadings matrix contains many zero entries, so that any single factor only impacts a relatively small number of elements of ${x}_i$. Hence, dimension reduction is often carried out through low rank and sparsity assumptions on the loadings matrix. See, for instance, \citet{West03, GhoshDunson09, bhattacharya11gammaprocessprior} and \citet{Ma13}. 

\begin{figure}
\captionsetup[subfigure]{justification=Centering}
\begin{subfigure}[t]{0.29\textwidth}
    \includegraphics[width=\textwidth]{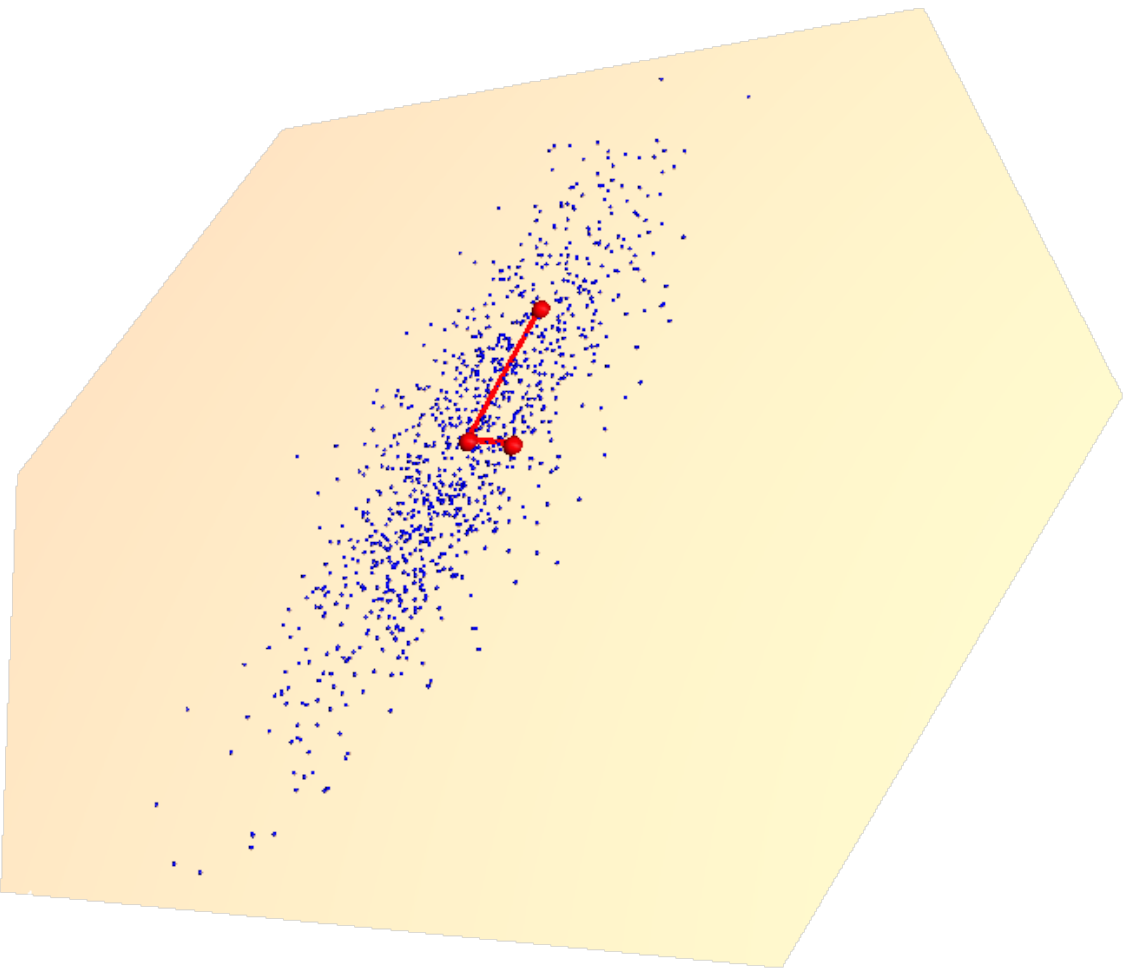}
    \caption{Data generated from $\text{N}_3({0}, \Lambda \Lambda^T)$}
    \label{fig:gauss_on_plane}
\end{subfigure}
\begin{subfigure}[t]{0.70\textwidth}
    \includegraphics[width = 0.45\textwidth]{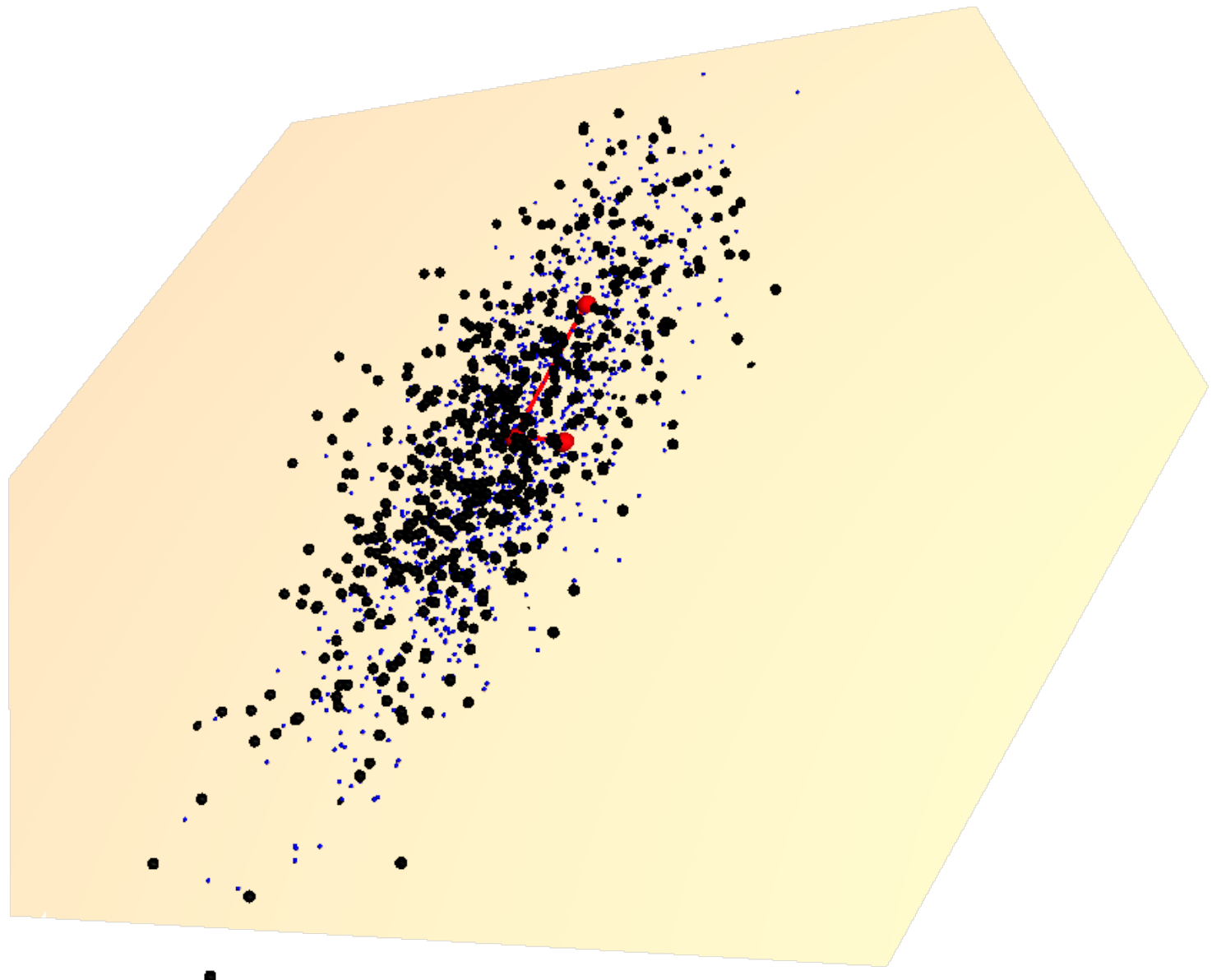}
      \includegraphics[width = 0.49\textwidth]{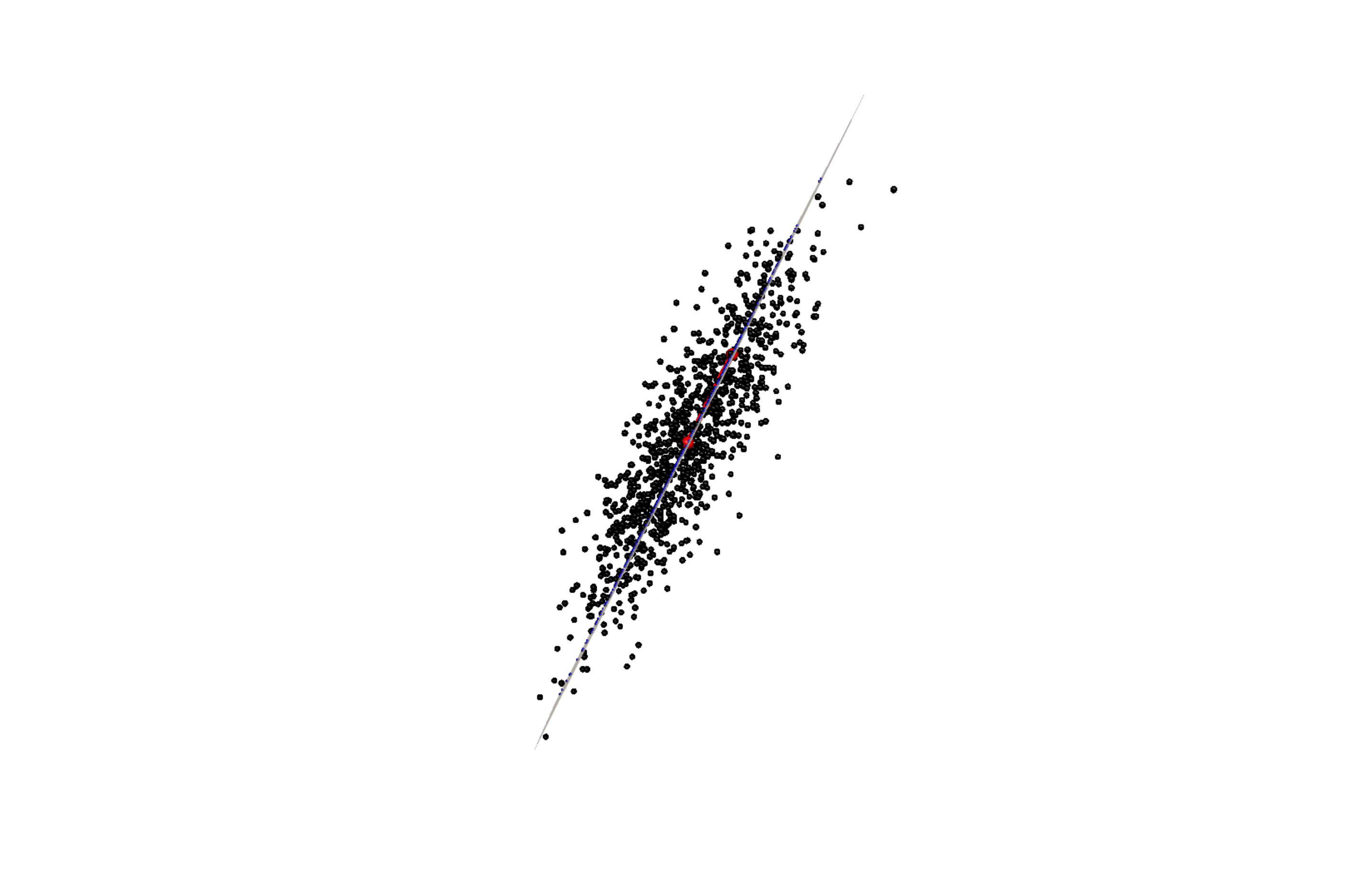}
       \caption{Data generated by adding random noise of distribution $\text{N}_3({0}, \Sigma)$ to the points in Fig.~\ref{fig:gauss_on_plane}}
        \label{fig:gauss_noisy}
\end{subfigure}
   \caption{Demonstration of the Gaussian linear factor model with $p = 3, k = 2$. The yellow plane represents the column space of $\Lambda$ and contains all of the blue points sampled from $\text{N}_3({0}, \Lambda \Lambda^T)$. The red vectors represent eigenvectors of $\Lambda \Lambda^T$; equivalently the left singular vectors of $\Lambda$. The figure on the right shows the view perpendicular to the plane, illustrating that the data have a Gaussian distribution centered around the plane.}
    \label{fig:demo_gauss}
\end{figure}

Our work begins from the observation that the multivariate Gaussian distribution cannot characterize nonlinear relationships in data. Consider the Gaussian linear factor model. To generate data from ${x}_i \sim \text{N}_p({0}, \Lambda \Lambda^T + \Sigma),$ we can first generate data from $\text{N}_p({0}, \Lambda \Lambda^T)$ and then add random $\text{N}_p({0}, \Sigma)$ noise. Figure \ref{fig:demo_gauss} shows a simulation example with $p = 3, k = 2.$ Figure \ref{fig:gauss_on_plane} shows data simulated from $\text{N}_3({0}, \Lambda \Lambda^T)$; since $\Lambda$ has rank 2, this is a degenerate Gaussian distribution on a linear subspace spanned by the columns of $\Lambda$. After adding Gaussian noise with covariance $\Sigma$ to the data in the plane, we obtain a distribution in $\mathbb{R}^3,$ as shown in Fig.~\ref{fig:gauss_noisy}. The Gaussian linear factor model assumes that the data are centered around a linear space spanned by the columns of $\Lambda$, and therefore cannot capture curvature.

\begin{figure}
   \captionsetup[subfigure]{justification=Centering}
\begin{subfigure}[t]{0.49\textwidth}
    \includegraphics[width=\textwidth]{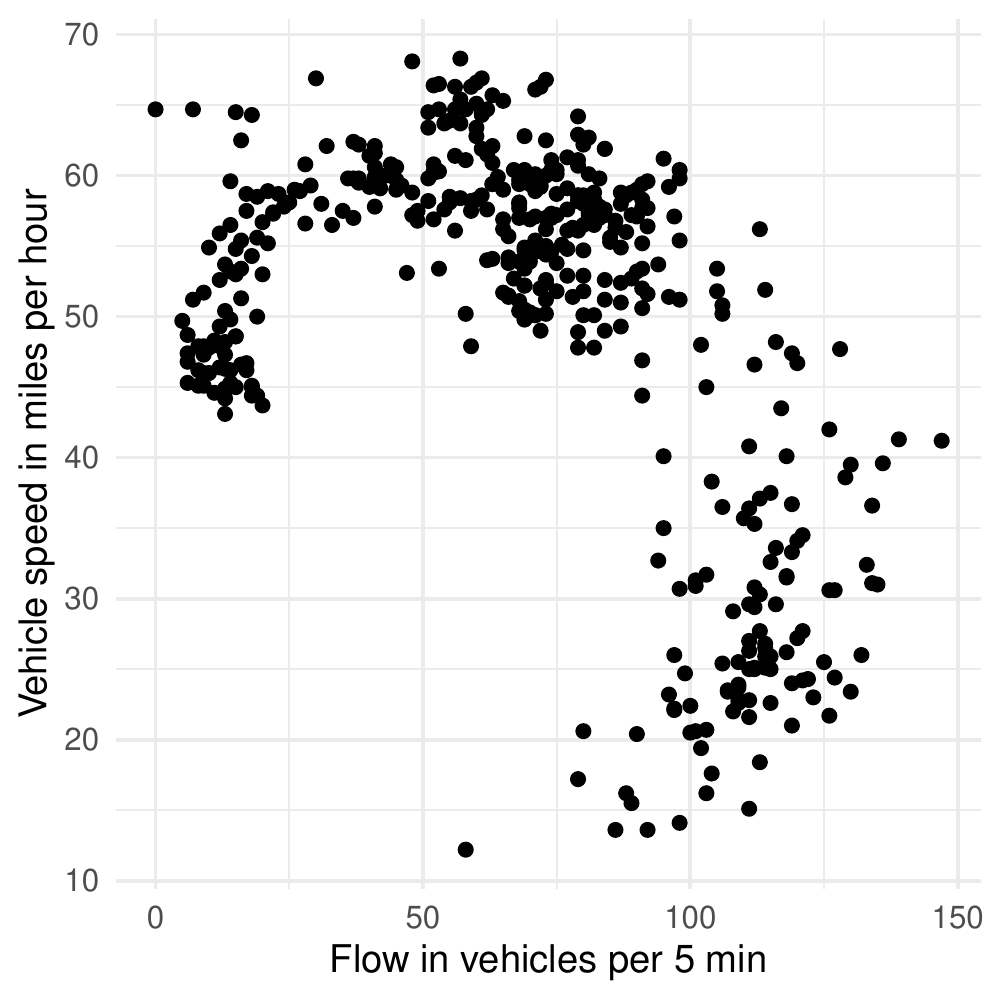}
    \caption{Scatter plot of California freeway data.}
    \label{fig:cal}
\end{subfigure}
\begin{subfigure}[t]{0.49\textwidth}
  \includegraphics[width=\textwidth]{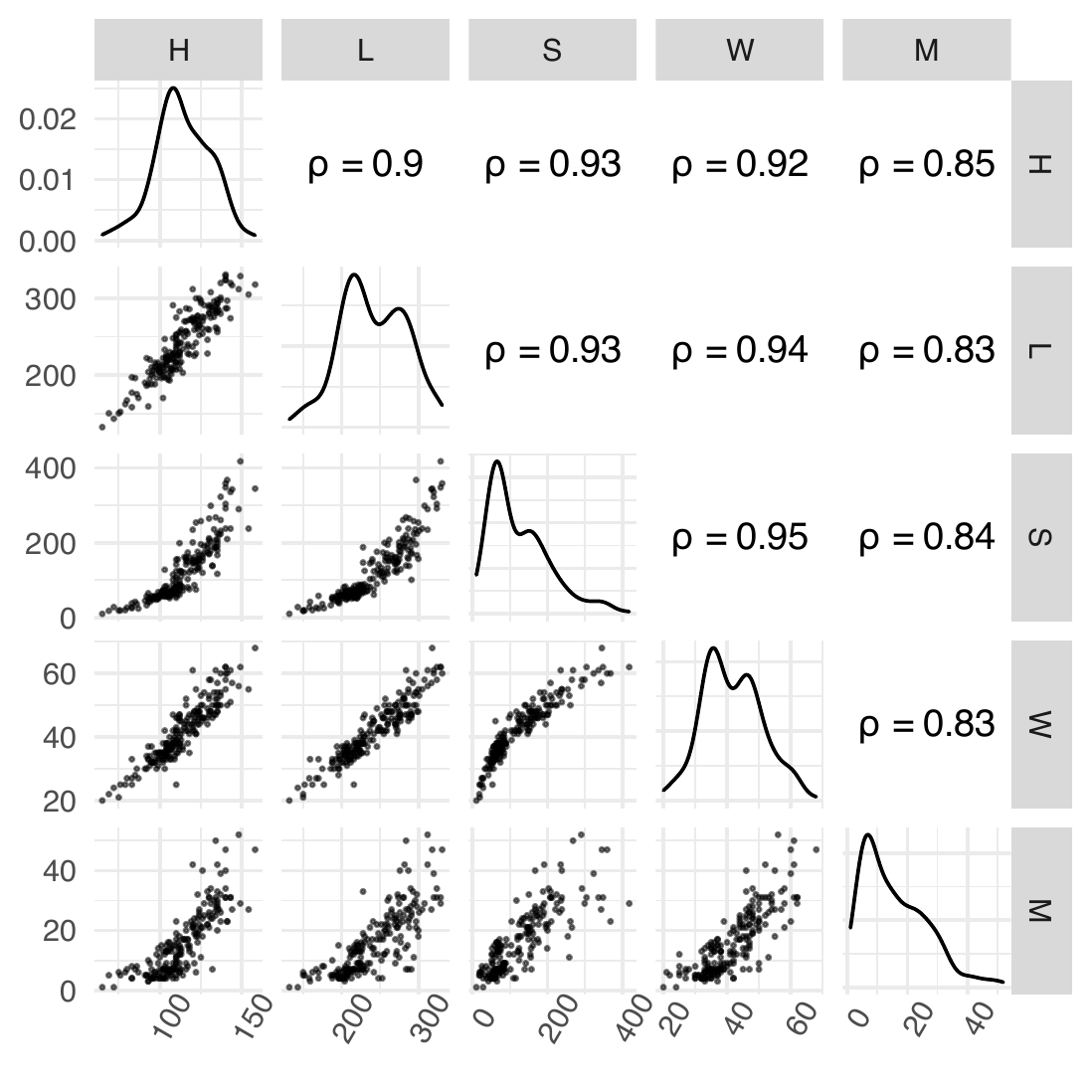}
    \caption{Scatter plot matrix of horse mussel data.}
    \label{fig:horsemussel}
\end{subfigure}
\caption{Scatter plots of two curved data sets. In particular, H, L, S, W, and M represent shell height (mm), shell length (mm), shell mass (g), shell width (mm) and muscle mass (g) respectively.\label{fig:demo_ex}}
\end{figure}

However, curved relationships are commonplace in real data.  We provide two motivating examples; one is a speed flow data set on a California freeway \citep{Einbeck11cal} and the other is an ecological data set of horse mussels in New Zealand \citep{CamdenMike1989Tdb}. Figure~\ref{fig:cal} shows that vehicle flow, per 5 minutes, and speed, in miles per hour, on the freeway have a curved relationship. Figure~\ref{fig:horsemussel} shows that body measurements, such as edible muscle mass (M) and shell width (W), of the horse mussels have curved dependence. 

There is a rich literature on nonlinear factor models that can capture curved dependence by replacing the linear loadings on $\eta_i$ with a more complex mapping. A general model is given by $x_i = g(\eta_i) + \epsilon_i$, with the same assumptions as a Gaussian linear factor model but a nonlinear function $g$ mapping $\mathbb{R}^k \to \mathbb{R}^p$. Popular nonparametric approaches include Gaussian process latent variables models, which give $g(\cdot)$ a Gaussian process prior \citep{TitLawrence10,LiChen16}, and variational auto encoders, which model $g(\cdot)$ using a deep neural network \citep{Kingma2013AutoEncodingVB, Rezende14vae}. Such methods are highly flexible but tend to be complex black boxes that have issues with reproducibility of results 
and non-identifiability, while being opaque to interpret. 

An alternative is to mix Gaussian linear factor models to capture nonlinear structure in the data by local linear Gaussian models; for example, refer to the rich literature on mixtures of factor analyzers \citep{Ghahramani97MFA, Mclachlan03MFA, Murphy20IMIFA}. However, in defining a factor model for every component of a mixture model, such models are heavily parameterized and can be difficult to fit reliably and interpret.  If there are not distinct clusters in the data but data have curved support, data will be broken up into many small clusters.

Other attempts to develop flexible parametric families of multivariate distributions have focused on capturing skewness and heavy tails in the data, among which the generalization of skew-elliptical distributions \citep{Azzalini19skewnormal,branco01skewelliptical} constitutes a prominent subset. Starting with multivariate skew-normal distributions \citep {azzalini96skewnormal}, there are extensions to multivariate skew-$t$ \citep{Gupta03skewt, kim03skewt, arevalillo15skewt}, scale mixtures of skew-normal \citep{kim08scalemixtureskewn, capitanio2012scalemixtureskewn}, skew-symmetric  \citep{jupp16skewsymmetric} and even broader families 
 \citep{Genton03generalizedskew-elliptical, azzalini03gskewelliptical,landsman17gskewelliptical}. An alternative extension of the skew-normal is the multivariate skew-slash distribution \citep{Wang06skewslash}. Certain of the above distributions can induce curvature, but in an indirect and hard to interpret manner.

Hence, there is a need to develop simple parametric models to characterize nonlinear dependence in data. In this article, we propose a class of parametric factor models that induce multivariate distributions supported near the surface of a hyper-ellipsoid. This Ellipsoid-Gaussian (EG) class is induced by using a Gaussian linear factor model with latent variables following a von Mises-Fisher distribution on a unit hyper-sphere. We show that the Ellipsoid-Gaussian distribution is surprisingly flexible in allowing curved relationships among variables and modeling of lower-dimensional structure, while including the Gaussian linear factor model as a special case.  All proofs are in the Supplementary Material \citep{song23eg_sm}.


We introduce the vMF linear factor model in Section \ref{sec:vmf_fac} and then integrate over the distribution of the latent factors to derive the Ellipsoid-Gaussian distribution and study its basic distributional properties in Section~\ref{sec:eg}. In Section \ref{sec:limit} we study the limiting behavior of the Ellipsoid-Gaussian distribution, making a connection between the vMF linear factor model and the Gaussian one. Section \ref{sec:computation} considers computational challenges, which motivate careful design of an appropriate MCMC algorithm. Sections \ref{sec:simulation} and \ref{sec:application} present a variety of simulation studies and real data applications to demonstrate the performance of the Ellipsoid-Gaussian distribution. Code is available as an R package on Github at \url{https://github.com/hanyu-song/ellipsoidgaussian}.

\section{The von Mises-Fisher linear factor model \label{sec:vmf_fac}}

To induce curved relationships among $x_{i1}, \ldots, x_{ip}$, a starting point is distributions on a sphere. Let $\mathcal{S}^{k-1}$ be a unit sphere in $\mathbb{R}^k$ centered at the origin and $\mathscr{S}^{k - 1}$ be the probability measure of the uniform distribution on $\mathcal{S}^{k -1}$. A simple density with respect to $\mathscr{S}^{k-1}$ for a random vector ${z} \in \mathcal{S}^{k-1}$ is
    $f({z}; \mu, \tau) = C_k(\tau) \exp\left(\tau \mu^T {z}\right),$
where $\mu$ is the mean direction with $\Vert \mu \Vert = 1$, $\tau \geq 0$ is a concentration parameter, $C_k(\tau) = ({\tau}/{2})^{k / 2 - 1}\{\Gamma(k / 2)I_{k /2 - 1}(\tau)\}^{-1}$ is the normalizing constant, and $I_{v}(\cdot)$ denotes the modified Bessel function of the first kind of order $v.$ This von Mises-Fisher density is symmetric about the mean direction $\mu$, with $\tau$ controlling concentration---a larger $\tau$ corresponds to higher concentration around $\mu.$ The inner product between two unit vectors $\mu^T {z}$ parametrizes the cosine distance between $\mu$ and ${z}$. This metric has been used in high-dimensional directional data clustering, such as for text and gene-expression data \citep{Banerjeeetal05, Reisingeretal10, GopalYang14}.

While the von Mises-Fisher distribution is not directly useful for the data ${x}_i$, since these data points are not exactly on a sphere, it can be used for latent factors in model~\eqref{equ:gaussfac} as follows:
\begin{align}
    {x}_i = {c} + {\Lambda} {\eta}_i + \epsilon_i, \quad \epsilon_i \sim \text{N}_p(0, \Sigma), \quad \eta_i \sim \text{vMF}(\mu, \tau),\quad i=1,\ldots,n.\label{equ:egdefinition}
\end{align}
To better understand this model, we first define some notation. Let the singular value decomposition of $\Lambda$ be ${U}_{p\times k}{S}_{k \times k}{V}_{k \times k}^T$, where ${U}^T{U} = {I}$, ${V}^T{V} = {I}$ and ${S}_{k \times k} = \diag\left(s_1, \ldots, s_k\right).$ The image of a unit sphere $\mathcal{S}^{k-1}$ under the affine transformation ${c} + {\Lambda}\mathcal{S}^{k-1}$ is a $k$-dimensional ellipsoid in $\mathbb{R}^p$; when $k < p$, the ellipsoid is degenerate. The center of the ellipsoid is ${c}$, its principal axes are represented by the left singular vectors ${U}$, and its semi-axes lengths by the singular values ${S}.$ The image of $\mathcal{S}^{k-1}$ after multiplication by $\Lambda$ can be understood as sequentially transforming the sphere by each matrix in the factorization: since the sphere is rotationally invariant, the image $\mathcal{S}^{k-1}$ under the rotation ${V}^T$ is still itself; ${S}$ stretches the sphere $\mathcal{S}^{k-1}$ into an ellipsoid centered at the origin with principal axes parallel to the coordinate axes and semi-axes lengths equal to the singular values $s_k$; finally, ${U}$ rotates the ellipsoid such that the principal axes are parallel to the column vectors of ${U}$ and embeds it into a $k$-dimensional linear subspace of $\mathbb{R}^p$. Adding ${c}$ translates the ellipsoid such that its center becomes ${c}.$ We then add Gaussian noise.

\begin{figure}
\captionsetup[subfigure]{justification=Centering}
\begin{subfigure}[t]{0.49\textwidth}
    \includegraphics[width=\textwidth]{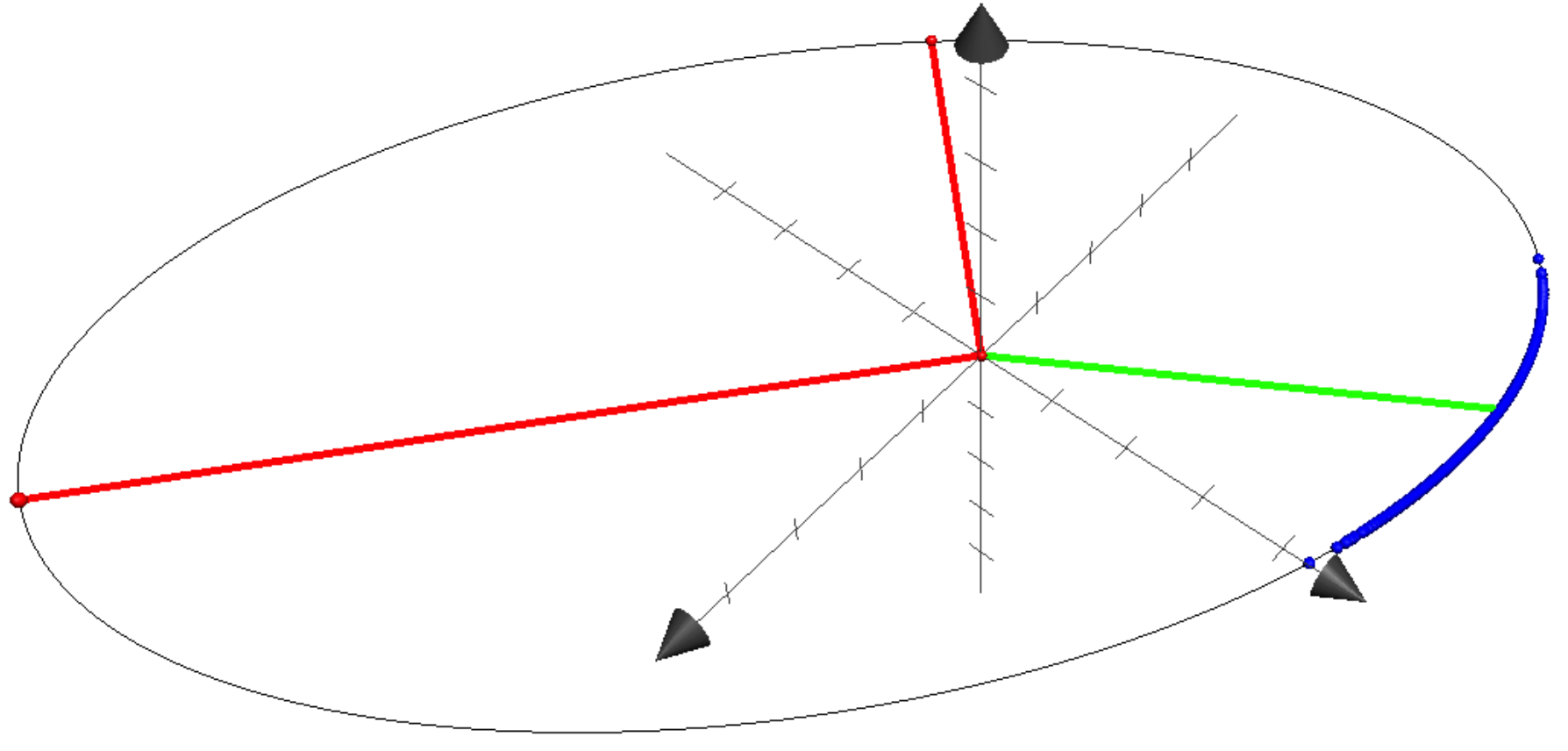}
    \caption{Data generated from a \emph{noiseless} von-Mises Fisher linear factor model with center ${c}$ at the origin.}
    \label{fig:ellipse}
\end{subfigure}
\begin{subfigure}[t]{0.49\textwidth}
    \includegraphics[width = \textwidth]{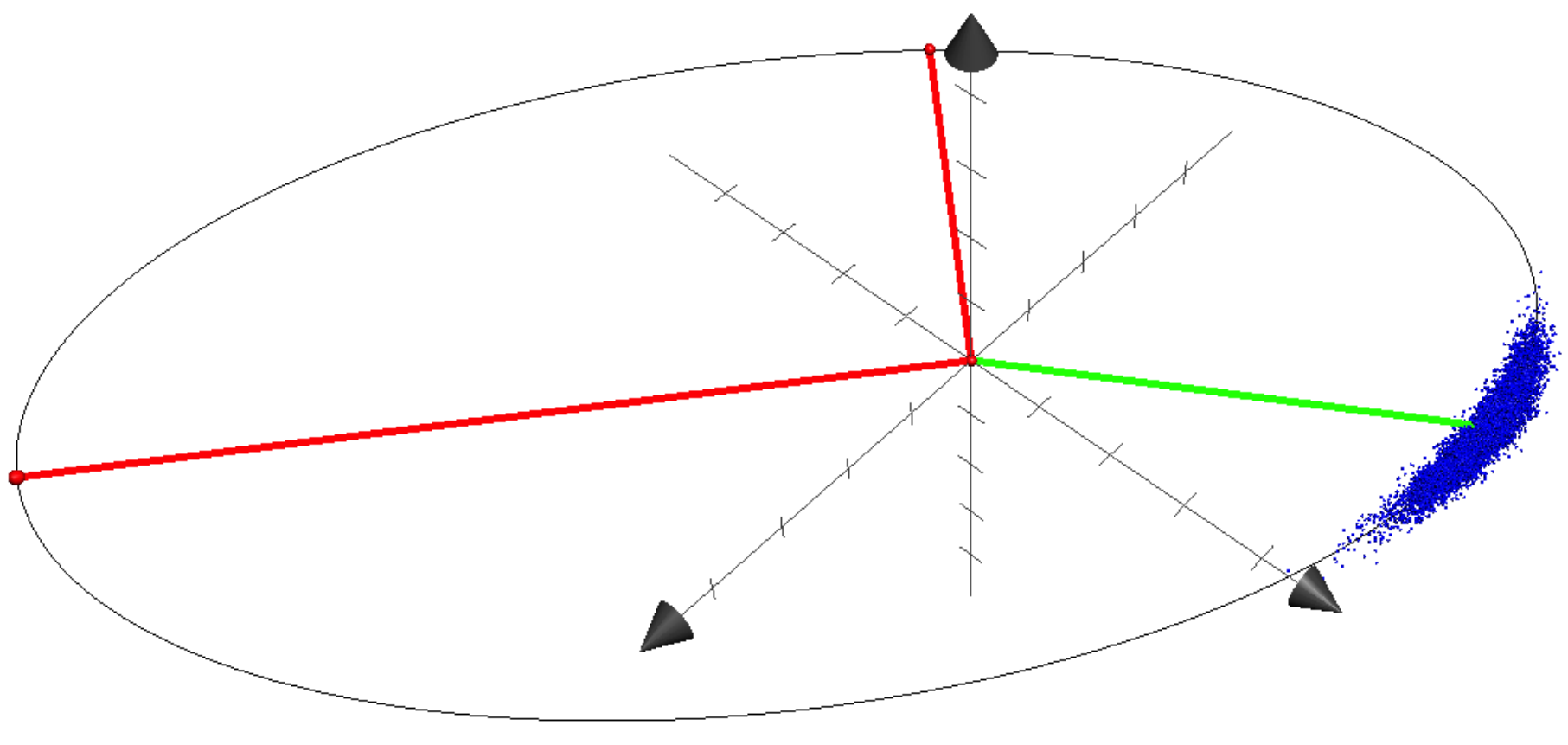}
       \caption{Data generated by adding random noise from $\text{N}_3({0}, \Sigma)$ to the points in Fig.~\ref{fig:ellipse}}
        \label{fig:ellipse_noisy}
\end{subfigure}
   \caption{Demonstration of the von Mises-Fisher linear factor model with $p = 3, k = 2.$ The green and red lines represent the mean direction $\Lambda \mu$ and principal axes of the ellipse, respectively. }
    \label{fig:demo_vmf}
\end{figure}

To visualize data generated from this model, consider an example with $p = 3, k = 2$. If we generate noiseless data from the model with $c = 0$, the points will lie on an ellipse as shown in Fig.~\ref{fig:ellipse}, with the center of the ellipse at the origin and the shape of the ellipse determined by $\Lambda$. The red line segments represent the principal axes of the ellipsoid and are also the left singular vectors of $\Lambda.$ We then add Gaussian noise around the points in Fig.~\ref{fig:ellipse} to obtain the desired distribution, as visualized in Fig.~\ref{fig:ellipse_noisy}, where the points are distributed around a curved surface, the ellipse, and this is how the curvature is captured in the model.

\begin{figure}
\captionsetup[subfigure]{justification=Centering}
\begin{subfigure}[t]{0.25\textwidth}
   \includegraphics[width = \textwidth]{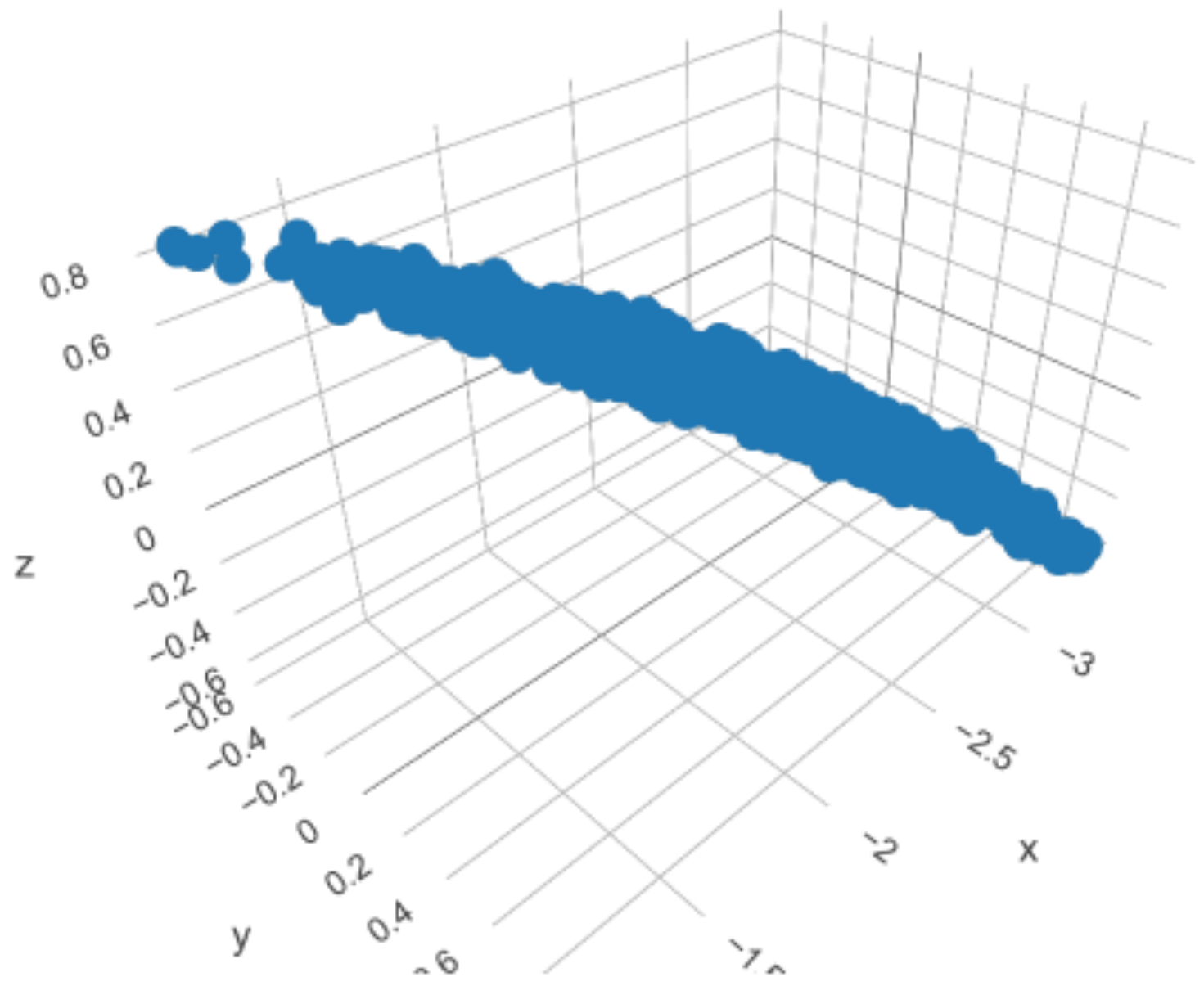}
   \caption{$\tau = 100$}
   \label{fig:eg_straight}
\end{subfigure}
  \begin{subfigure}[t]{0.26\textwidth}
  \includegraphics[width = \textwidth]{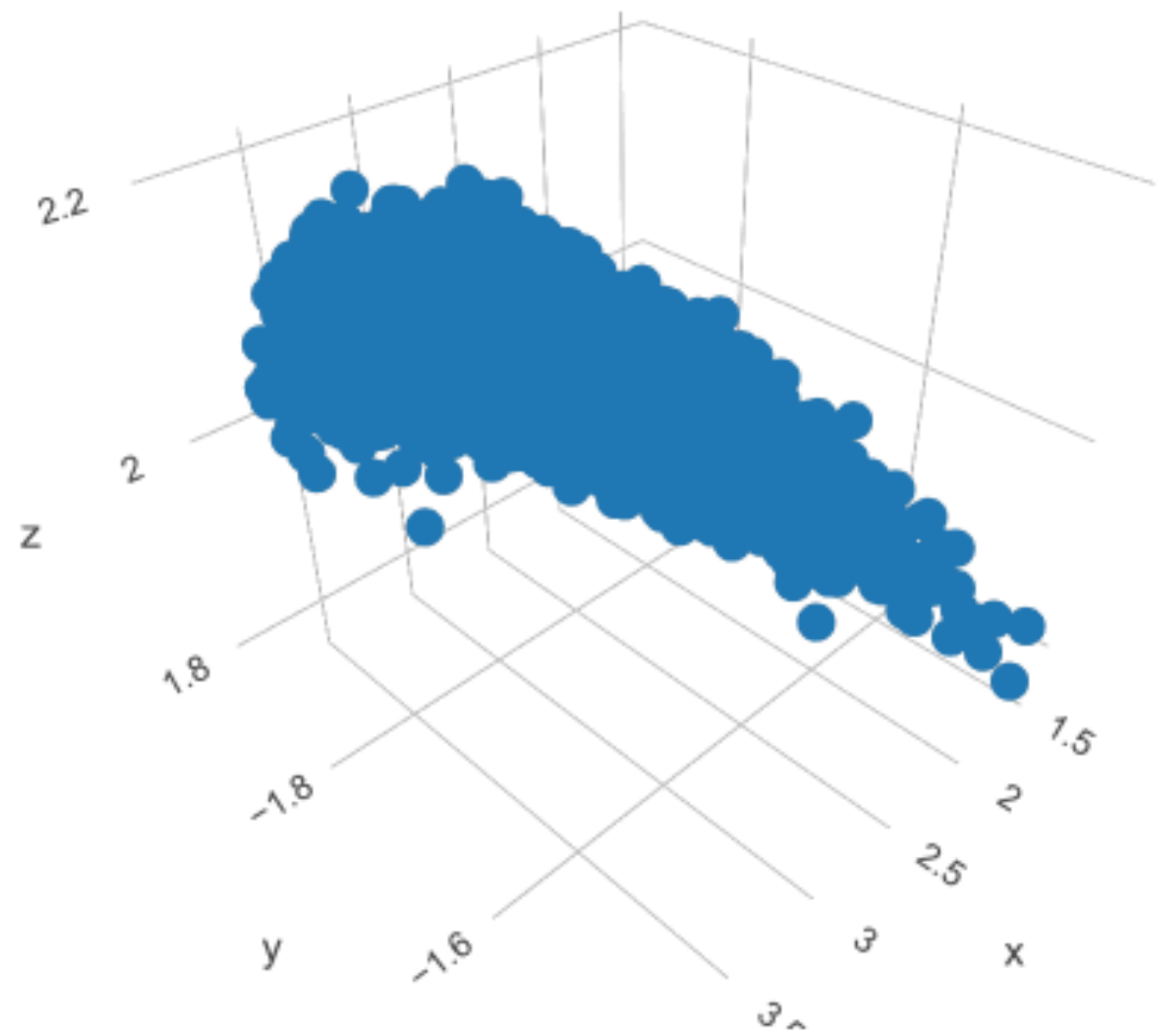}
    \caption{$\tau = 100$}
    \label{fig:eg_banana2}
  \end{subfigure}
  \begin{subfigure}[t]{0.25\textwidth}
      \includegraphics[width =\textwidth]{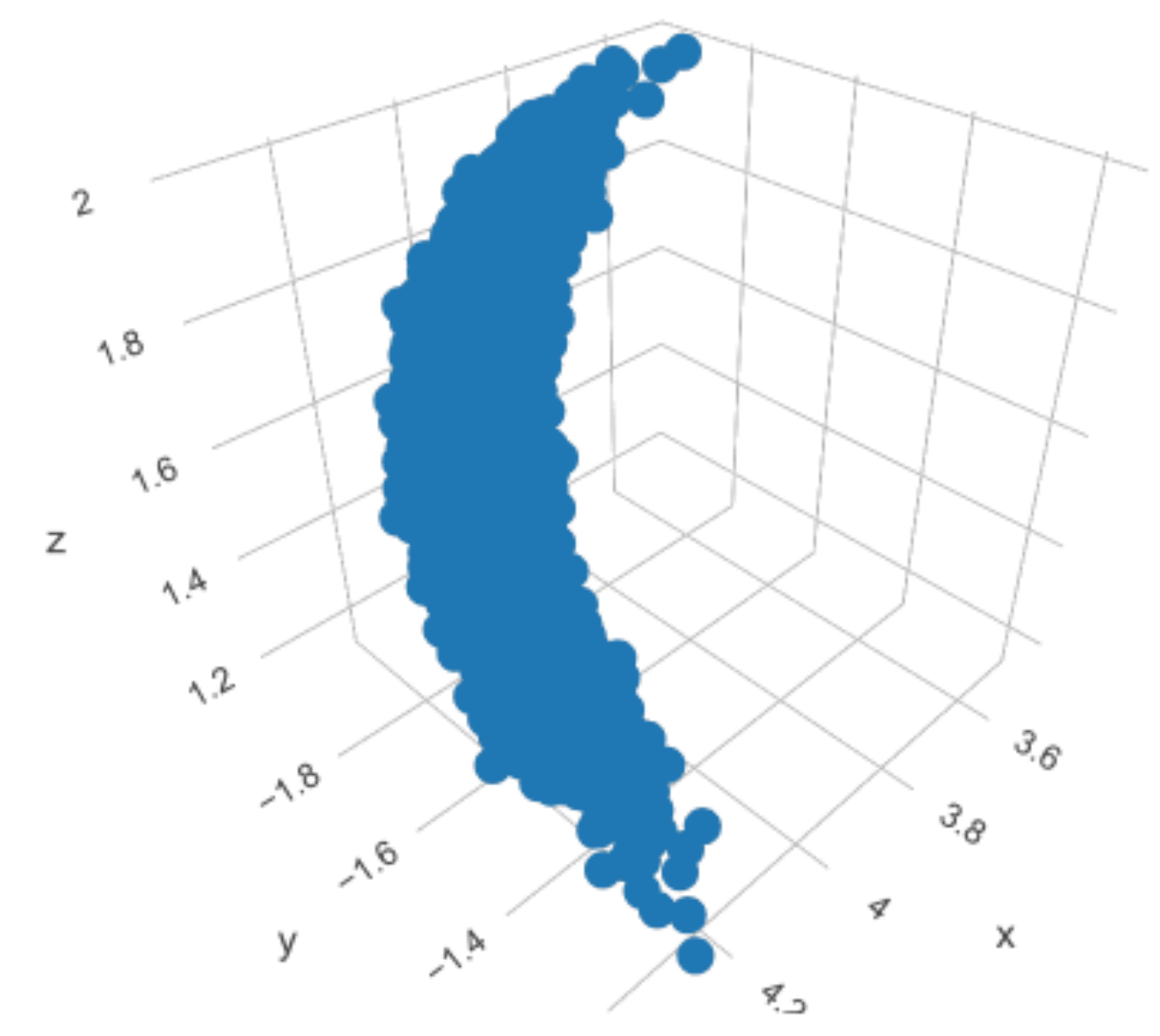}
      \caption{$\tau = 100$}
      \label{fig:eg_symmetric}
  \end{subfigure}
  \begin{subfigure}
      [t]{0.25\textwidth}
      \includegraphics[width =\textwidth]{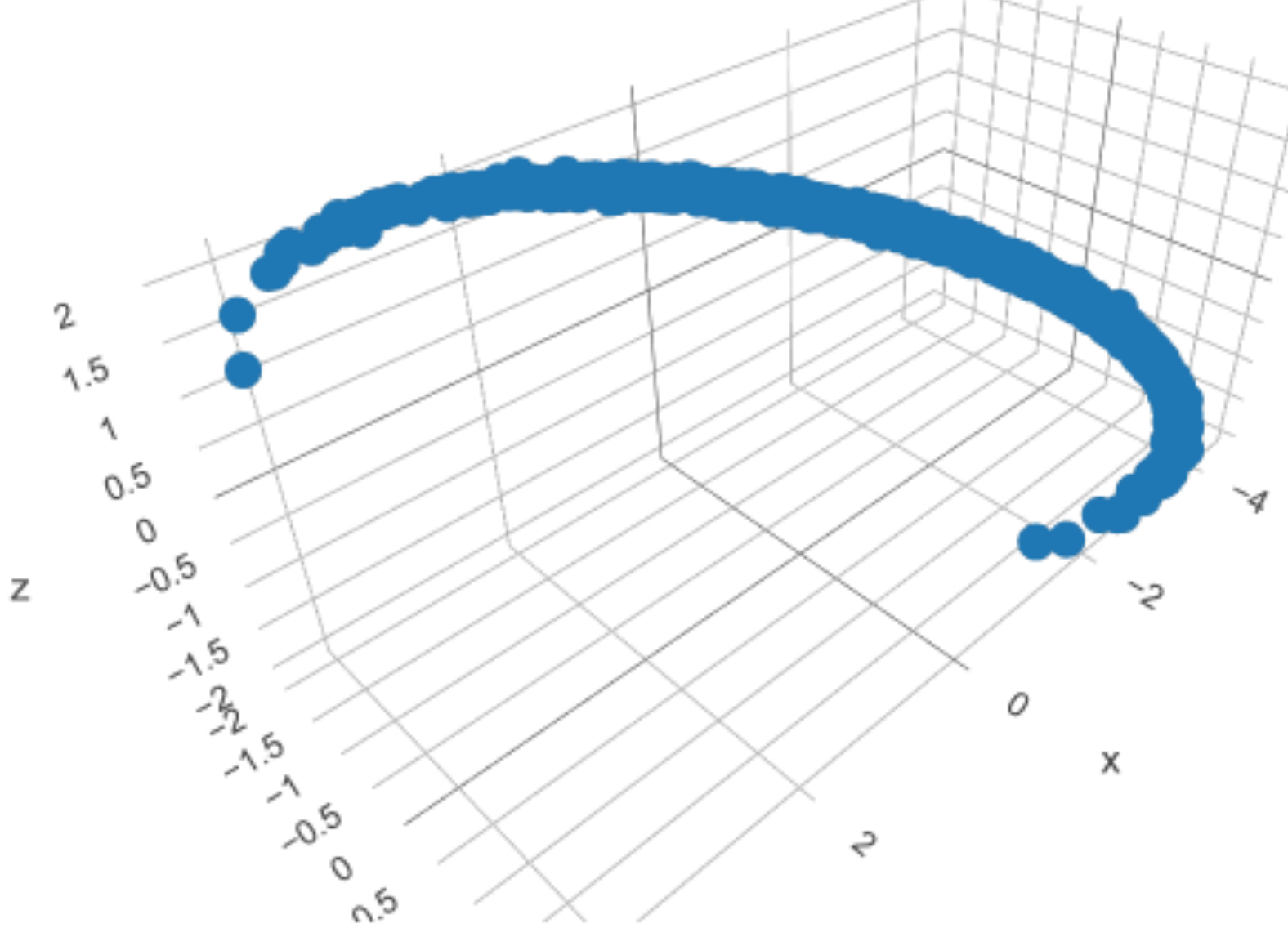}
      \caption{$\tau = 5$}
      \label{fig:eg_wrapped}
  \end{subfigure}
   \begin{subfigure}
      [t]{0.25\textwidth}
      \includegraphics[width =\textwidth]{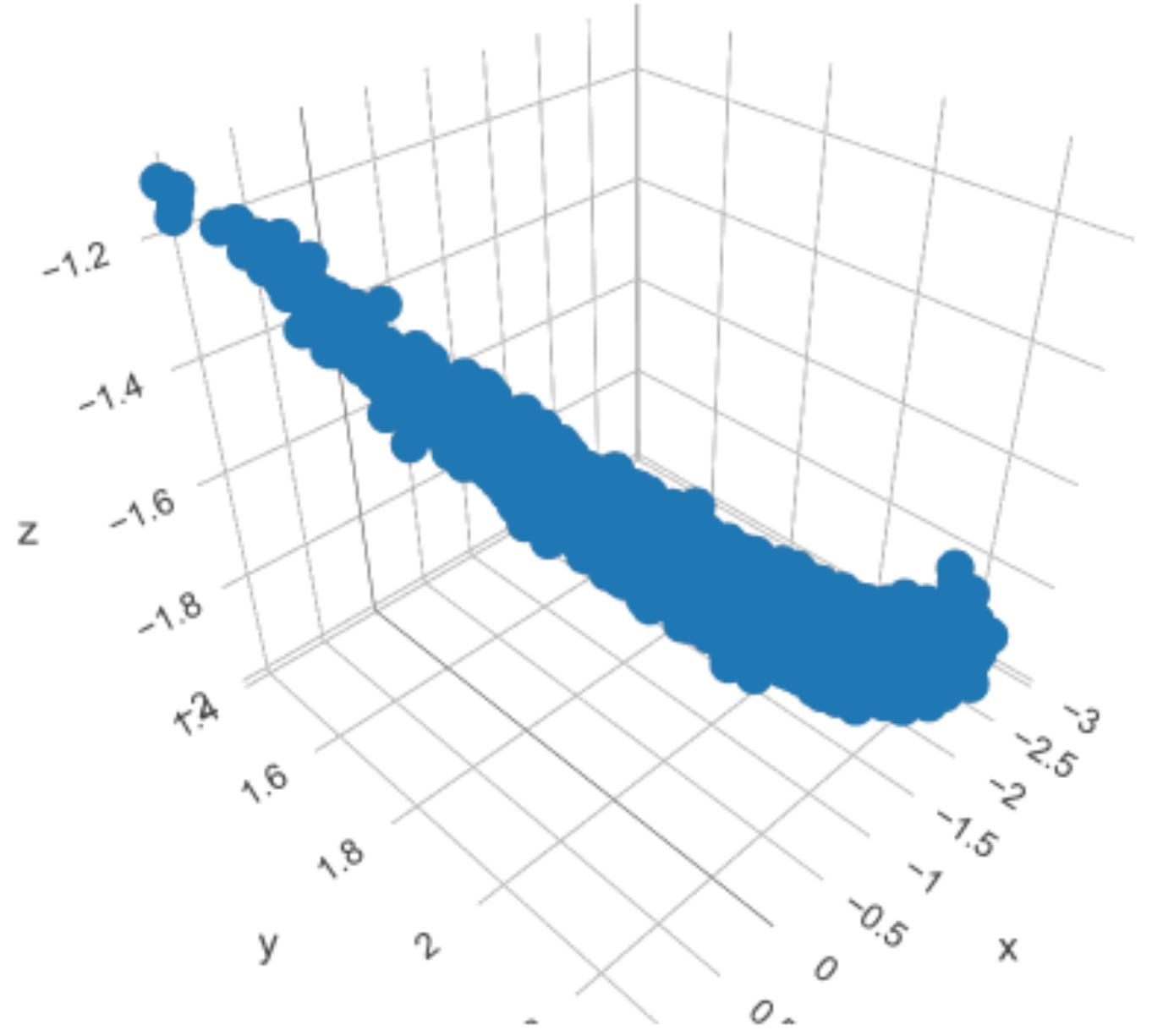}
      \caption{$\tau = 50$ \label{fig:eg_asymmetric2}}
  \end{subfigure}
   \begin{subfigure}
      [t]{0.25\textwidth}
      \includegraphics[width =\textwidth]{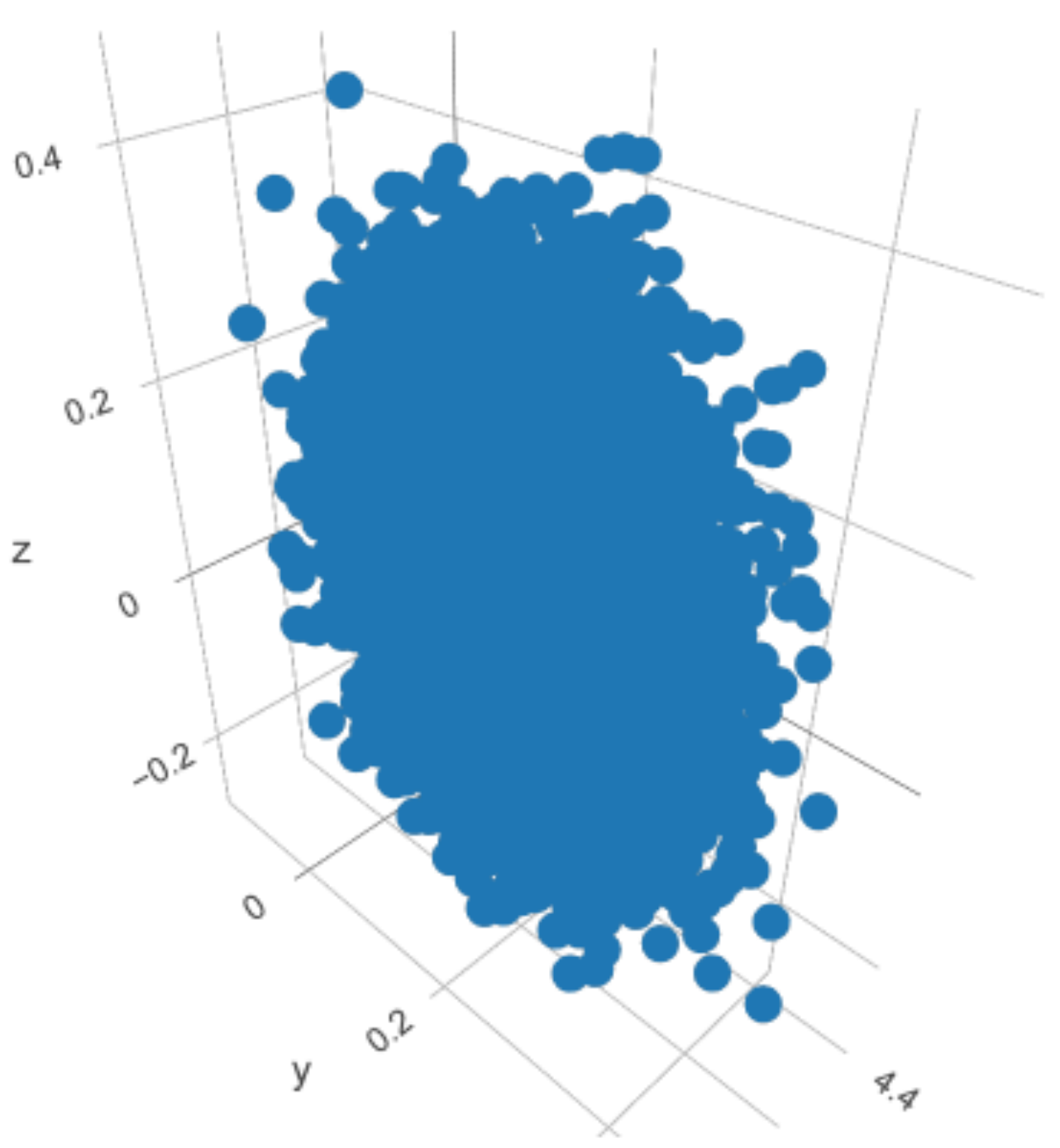}
      \caption{$\tau = 10^4$   \label{fig:eg_gaussian}}
  \end{subfigure}
   \caption{Scatter plots of data sampled from the Ellipsoid-Gaussian in $\mathbb{R}^3$ with varying parameters.\label{fig:egscatterplot}}
\end{figure}

Figure~\ref{fig:egscatterplot} shows more simulated examples from this model. The various shapes of the data clouds suggest tremendous flexibility of the model in accommodating different data --- data that exhibit symmetry, asymmetry, uneven curvature or data that are close to Gaussian. 

    \section{Ellipsoid-Gaussian distribution \label{sec:eg}}
Analogous to the Gaussian latent factor model, we wish to marginalize over the distribution of the latent factor $\eta$ to obtain the marginal distribution of ${x}$; we call the resulting distribution \emph{Ellipsoid-Gaussian}.  The density of ${x}$ can be calculated in closed-form, and ${x}$ is supported on the entirety of $\mathbb{R}^p.$   In order to state the density, we recall the Fisher-Bingham distribution, which is a distribution on $\mathcal{S}^{k-1}$ that includes the von Mises-Fisher distribution as a special case. A random vector ${y}$ with a Fisher-Bingham distribution has density with respect to $\mathscr{S}^{k-1}$:
\begin{align*}
    f({y}; \kappa, \vartheta, A) = \frac{1}{\varsigma(\kappa \vartheta, A)} \exp\left(\kappa \vartheta^T{y} - {y}^T A {y}\right),
\end{align*}
where ${y}, \vartheta \in \mathcal{S}^{k -1}$, $\kappa \geq 0$, $A \in \mathbb{R}^{p \times p}$ is a symmetric matrix, and $\varsigma(\kappa \vartheta, A)$ is a normalizing constant that
is commonly estimated via a saddlepoint approximation \citep{KumeWood05} which is accurate, fast and numerically stable. The Fisher-Bingham distribution has been studied in \citet{Wood88FB,Hoff09FB,kent2013new} and many other works.  

Proposition~\ref{prop:density} shows the marginal density of the Ellipsoid-Gaussian distribution. 
\begin{prop}[Properties of Ellipsoid-Gaussian]\label{prop:density} 
Assume ${x} = {c} + \Lambda \eta + \epsilon,\quad \epsilon \sim \text{N}_p({0}, \Sigma), \quad \eta \sim \text{vMF}(\mu, \tau),$ and $\Sigma = \diag\left(\sigma_1^2, \ldots, \sigma_p^2\right)$.
\begin{enumerate}
\item  Marginalizing over the distribution of $\eta$ yields density  
\begin{align*}
f_\text{EG}(x) &= \frac{C_k(\tau)}{(2\pi)^{\frac{p}{2}}\prod_{i = 1}^p \sigma_i} \exp\left\{-\frac{1}{2} (x - {c})^T \Sigma^{-1} (x - {c})\right\}\\&
   \varsigma\left\{ \tau \mu + \Lambda^T \Sigma^{-1
}(x - {c}) , \frac{\Lambda^T \Sigma^{-1}\Lambda}{2}\right\},
\end{align*}
with respect to the Lebesgue measure on $\mathbb{R}^p,$ where $\varsigma(\kappa \vartheta, A)$ is the normalizing constant in a Fisher-Bingham density. 
\item The Fisher-Gaussian distribution \citep{MukLiDunson19} is a special case of the Ellipsoid-Gaussian distribution with $\Lambda = r{I}_p$ and $\Sigma = \sigma^2{I}_p.$
\item The marginal distribution of any sub-vector of $x$ also follows an Ellipsoid-Gaussian distribution. Specifically, suppose the index set of the random vector is $I \subset [p]$ and let subscript $I$ represent elements from the rows of a matrix or elements from a vector, whose index is in set $I.$ Then $x_I$ follows a Ellipsoid-Gaussian with density
\begin{align*}
    f_\text{EG}(x_I) &= \frac{C_k(\tau)}{(2\pi)^{\frac{\mid I\mid }{2}}\prod_{i \in I} \sigma_i} \exp\left\{-\frac{1}{2} (x_I- {c_I})^T \Sigma_I^{-1} (x_I - {c_I})\right\}\\&
   \varsigma\left\{ \tau \mu + \Lambda_I^T \Sigma_I^{-1
}(x_I - {c}_I) , \frac{\Lambda_I^T \Sigma_I^{-1}\Lambda_I}{2}\right\},
\end{align*}
where $\Sigma_I = \diag(\sigma_j, j \in I).$
\item The expectation of ${x}$ is ${c} + \rho_k(\tau) \Lambda \mu,$
where $\rho_k(\tau) = I_{k / 2}(\tau)/I_{k / 2 - 1}(\tau).$ \label{equ:mean_EG}
\item The covariance of ${x}$ is ${\rho_k(\tau)}/{\tau}\Lambda \Lambda^T +\left\{1 -  \frac{k}{\tau}\rho_k(\tau) - \rho_k^2(\tau)\right\}\Lambda \mu (\Lambda \mu)^T + \Sigma.$ \label{equ:sec_mom_EG}
\end{enumerate}
\end{prop}

This Proposition has several interesting ramifications. The term $\varsigma$ is not part of the normalising constant as it depends on the random variable ${x}$; this leads to some challenges in model fitting.
The Fisher-Gaussian distribution in \citet{MukLiDunson19} is a special case, which corresponds to the Ellipsoid-Gaussian with no dimension reduction ($p = k$), spherical Gaussian noise, and $\Lambda$ being a scalar multiple of the identity matrix. In addition, unlike the Gaussian linear factor model, ${c}$ is no longer the expectation of ${x},$ suggesting that we cannot simply center the data prior to analysis and remove ${c}.$ 
\subsection{Identifiability of the model parameters \label{sec:identify}}

\begin{figure}
\centering
	\includegraphics[width = 0.35\textwidth]{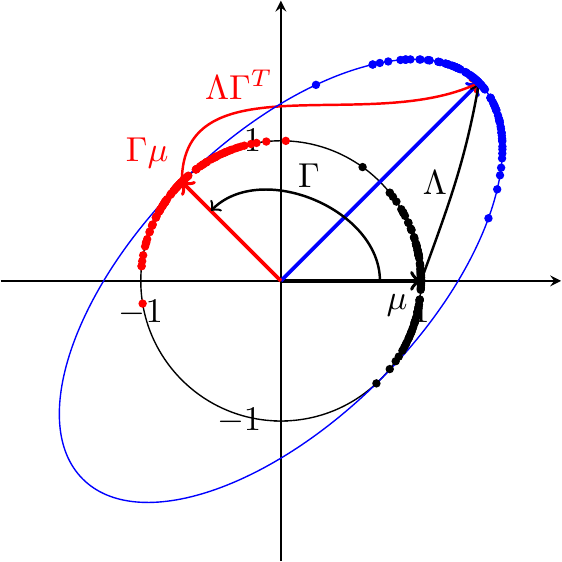}
\caption{Rotating the samples from a vMF$(\mu, \tau)$ (black) by $\Gamma$ results in samples from a vMF$(\Gamma \mu,\tau)$ (red). Mapping the black points by $\Lambda$ and the red points by $\Lambda \Gamma^T$, respectively, results in the same set of points (blue).}
\label{fig:vmf_rotate}
\end{figure}

An Ellipsoid-Gaussian distribution contains parameters $\{{c}, {\Lambda}, {\mu}, \tau, \Sigma\}$ of total dimension $(p + 1)(k + 1) + p.$ In this section, we show that some of these parameters are not identifiable.
\begin{prop}
\label{prop:rotationinvariant}
In model \eqref{equ:egdefinition}, $\Lambda$ is only identifiable up to orthogonal transformation. 
\end{prop}
Figure \ref{fig:vmf_rotate} gives a visual representation of this proposition, which shows that the nonidentifiability of $\Lambda$ is due to a transformation property of the von Mises-Fisher distribution. 
Model \eqref{equ:gaussfac} shares this property, due to rotational invariance of the distribution of 
 the latent factors $\text{N}{({0}, {I}_k)}$, with the image of $\eta$ under an orthogonal transformation $\Gamma$ still being $\text{N}{({0}, {I}_k)}$. 
In \eqref{equ:egdefinition}, however, an orthogonal transformation $\Gamma$ of $\eta$ results in a different von Mises-Fisher distribution, having the same concentration parameter $\tau$ but with a different mean vector $\Gamma \mu.$ Therefore, a rotation of $\eta$ necessarily results in a rotation in $\Lambda$ as well.
The following proposition provides further insights.

\begin{prop} \label{prop:egMGF}
The moment generating function of the Ellipsoid-Gaussian distribution is
\begin{align*}
    M_\text{EG}(t) &= \exp\left(t^T{c} + \frac{1}{2}t^T\Sigma t\right)\frac{C_k(\tau)}{C_k(\Vert \Lambda^Tt + \tau\mu \Vert)}.
\end{align*}
\end{prop} This proposition suggests a different parameterization. Let 
\begin{align*}
    \Vert \Lambda^T t + \tau \mu \Vert &= \sqrt{t^T\Lambda \Lambda^Tt + 2\tau t^T \Lambda\mu  + \tau^2}.
\end{align*}
Hence, the moment generating function depends on $\Lambda$ and $\mu$ only through $\Lambda\Lambda^T$ and $\Lambda\mu$. By Proposition \ref{prop:density}, $\Lambda\mu$ is the mean direction if the underlying ellipsoid is centered at the origin. The symmetric matrix $\Lambda\Lambda^T$ with spectral decomposition ${US}^2{U}^T$ defines a $k$-dimensional ellipsoid consisting of points satisfying ${x}^T{U}{S}^{-2}{U}^T{x} = 1$. Instead of using $\Lambda$ and $\mu$, we can view the distribution as being parameterized by the supporting ellipsoid and the mean direction $\Lambda \mu$.

\subsection{Limiting behavior\label{sec:limit}}
In this section, we show that the Gaussian linear factor model is a limiting case of the Ellipsoid-Gaussian distribution. The following Lemma states that, as the concentration of a von Mises-Fisher distribution increases, the distribution approaches a degenerate Gaussian; a visualization of this lemma is included in the Supplementary Material \citep{song23eg_sm} along with its proof.
\begin{lemma}[Limiting behavior of von Mises-Fisher] \label{lem:vmf_limit}
 Assume that $\eta \sim \text{vMF}(\mu, \tau)$. As $\tau \to \infty$, the quantity $\tau^{1/2}(\eta - \mu)$ converges in distribution to a multivariate Gaussian with mean ${0}$ and covariance matrix ${I}_k - \mu \mu^T$, which is supported on the hyperplane perpendicular to $\mu.$
\end{lemma}

Using Lemma~\ref{lem:vmf_limit}, we are able to show that as $\tau$ goes to infinity, \eqref{equ:egdefinition} reduces to a Gaussian linear factor model with $(k-1)$-dimensional latent factors in $\mathbb{R}^{k-1}.$


\begin{prop}[Limiting behavior of Ellipsoid-Gaussian] \label{prop:vmf_to_normal}
Suppose we have a hyper-ellipsoid with axes represented by unit vectors ${u}_1, \ldots, {u}_k$ and semi-axis lengths $s_1 \geq s_2 \dots \geq s_k \geq 0$, with ${U} = [{u}_1 \ldots {u}_k]$ and ${S} = \text{diag}(s_1,\ldots, s_k)$. Suppose ${x}_\tau$ follows \eqref{equ:egdefinition}, with $\Lambda^{(\tau)} = {U}{S}^{(\tau)},$ where $S^{(\tau)} = \sqrt{\tau}\text{diag}(s_1,\ldots,s_{k-1},0)$ and latent factors $\eta_\tau \sim \text{vMF}(e_k, \tau)$ around $\mu = e_k$. Then ${x}_\tau$ converges in distribution as $\tau \to \infty$ to a Gaussian latent factor model with $k-1$ latent factors and loadings matrix ${U}_{-k}{S}_{-k}$ where ${U}_{-k} = [{u}_1 \ldots {u}_{k-1}]$ and ${S}_{-k} = \text{diag}(s_1,\ldots, s_{k-1})$, namely ${x}_{\infty} = {c} + \tilde{\Lambda} \tilde{\eta} + \epsilon,$ where $\tilde{\eta} \sim \text{N}({0}, {I}_{k-1}),$ $\tilde{\Lambda} = {U}_{-k}{S}_{-k}.$
\end{prop}

\section{Posterior computation\label{sec:computation}}
\subsection{Sampling algorithm}
Our extensive experiments have shown that standard Markov chain Monte Carlo algorithms, such as Gibbs sampling and Hamiltonian Monte Carlo, fail to perform adequately for posterior sampling in models involving Ellipsoid-Gaussian likelihoods. In this section, we describe the computational challenges and motivate the algorithm we choose---a hybrid of geodesic stochastic Nosé-Hoover thermostat \citep{liu16sgMCMC} and adaptive Metropolis \citep{vihola12AM}. The detailed sampling procedures, including an extensive discussion of prior specification, gradient computation and an outline of the algorithm, can be found in the Supplementary Material \citep{song23eg_sm}.

In general, samplers which rely on instantiating the latent variables $\{\eta_i \}$ tend to be subject to poor mixing, including Gibbs samplers.  This is due to well known problems with Gibbs sampling in latent factor models, which motivated pseudo marginal algorithms \citep{Andrieu09pseudo,Beaumont03pseudomarg, vihola12AM}.  We can bypass the need for pseudo marginal algorithms, which have also had poor performance in our experiments, due to the availability of a closed form likelihood marginalizing out the latent variables.

Marginalizing out the latent factors and using the resulting Ellipsoid-Gaussian likelihood in posterior sampling brings challenges. First, the term $\xi\{\tau \mu + \Lambda^T\Sigma^{-1}(x-{c}), {\Lambda^T \Sigma^{-1} \Lambda}/{2}\}$ needs to be approximated
 \citep{KumeWood05}, but repeatedly applying such
  approximations becomes slow.  
Second, the gradient tends to change drastically with small changes in parameters, making it challenging to define efficient proposals.  One failed example is the 
 Barker proposal \citep{livingstone2020barker}. Even when proposal values are somewhat close to the current samples, the drastic changes in the gradient result in acceptance probability close to zero; to improve acceptance the algorithm adapts to propose tiny changes. The recently proposed transport Markov chain Monte Carlo \citep{duan2021transport} failed for similar reasons.  Third, Gibbs-type updates lead to poor mixing, even with the latent factors marginalized out.  For example, when we condition on the center ${c}$ and the shape of the ellipsoid as determined by $\Lambda$, there is often not much uncertainty in $\mu$, a parameter related to the mean direction of the data. As a result the chain tends to get stuck in a local mode.  
Fourth, $\mu$ is constrained to be on a sphere. A typical algorithm would require transforming $\mu$ to an unconstrained space $\mathbb{R}^k.$ Specifically, to update $\mu$, we update an unconstrained vector $\tilde{\mu}$ and map it back to $\mu$ with  $\tilde{\mu} / \Vert \tilde{\mu} \Vert_2.$ $\tilde{\mu}$, unfortunately, tends to move in the direction close to being perpendicular to the unit sphere centered at the origin, which translates into minimal movement in $\mu$, as our experiments show. In addition, the update of the loadings matrix is also not very efficient, even with the use of popular shrinkage priors, such as the Dirichlet-Laplace \citep{Bhattacharya15DL}. This makes us wonder whether directly updating the axes directions and lengths can lead to faster convergence since the dimension of the parameter space would be drastically reduced.

A possible remedy to the first issue is to use stochastic gradient algorithms; a small subset of the data are involved in a given iteration, leading to dramatic speedup.  By relying on a small step size, one can avoid the need for a Metropolis-Hastings correction \citep{Ma15sgMCMC}, which circumvents the second issue. Also, by updating all the parameters in a single block, we address the third challenge. Our solution to the fourth problem is inspired by sampling schemes designed for distributions on manifolds embedded in Euclidean space with a known geodesic flow; see \citet{byrne13gmc} and \citet{liu16sgMCMC} for more details.

Based on the above considerations, and on thorough experiments trying different algorithms, we use 
geodesic stochastic gradient Nosé-Hoover thermostats \citep{liu16sgMCMC}; we choose this dynamics because of the ease of tuning and robust performance. One weakness, however, is that all parameters share a single step size but the variances of their associated gradient are on vastly different scales. A discussion of our failed attempts to resolve this issue is included in the Supplementary Material \citep{song23eg_sm}. Our solution is to add an additional transition kernel for the parameters that would benefit from a larger step size, such as $\Sigma$ and $\tau;$ robust adaptive Metropolis  \citep{vihola12AM} provides an effective way to define this transition kernel. 
\subsection{Estimating the parameter \texorpdfstring{$\mathbf{c}$}{c}}
The parameter $\mathbf{c}$ is rarely the inferential target. In Gaussian linear factor models, $\mathbf{c}$ is typically estimated by the data mean $\bar{X},$ since it represents both the maximum likelihood and the first-moment estimator. As this is not the case for Ellipsoid-Gaussian, we use the property that $\mathbf{c}$ represents the center of the underlying ellipsoid to estimate $\mathbf{c}$. Cayley transform ellipsoid fitting (CTEF) \citep{melikechi2023ellipsoid}, a state-of-the-art ellipsoid fitting algorithm, best suits our need. Data in most applications are noisy and do not concentrate around a \emph{whole} ellipsoid, and CTEF often performs well even in these cases. With a quality estimate of $\mathbf{c},$ we can either fix or update $\mathbf{c}$ through the sampling process; our flexible sampler allows both options.

\section{Simulation studies\label{sec:simulation}}
In this section we perform simulation studies to compare the Ellipsoid-Gaussian to other models that might be used to fit data with curvature.
\subsection{Experiment setups}
We compare to three methods to illustrate our ability to characterize multivariate distributions with curved dependence. The first method is infinite mixtures of infinite factor analyzers \citep{Murphy20IMIFA}, abbreviated as mixtures of factor analyzers, and is implemented by the R package \citep{Murphy21Rimifa} (version 2.1.10) \texttt{IMIFA}. The second method is infinite Gaussian linear factor models, abbreviated as GLF, and is implemented by the R package  \texttt{infinitefactor}\citep{poworoznek20Rfac} (version 1.0). The third method is Bayesian Gaussian copula factor models, abbreviated as bfa, and is implemented by the R package \texttt{bfa} (version 1.4). Our method is implemented by the R package \texttt{ellipsoidgaussian} (version 1.0) \citep{song23eg}. All four packages use Markov chain Monte Carlo. The experiments are run on a iMac with 16G of memory and an M1 chip.

For each simulated data setting in Section~\ref{sec:simulation_setting} we generate 10 replicates of data sets of sizes $\{100,  \cdots, 1000\}$ and  test set of size 1000. We evaluate performance on the test set in three different ways. First, we estimate the log posterior predictive density as  $$h_q(X_{\text{test}}) := \sum_{i \in \text{test set}} M^{-1}\sum_{m = 1}^M \log q(x_i; \theta^{(m)}),$$ where $q(x_i; \theta^{(m)})$ is the likelihood of $x_i$ evaluated at the $m$th draw $\theta^{(m)}$ from the posterior. 
We do not calculate $h_q(X_{test})$ for the copula factor model as it lacks a complete likelihood specification.
The second basis for comparison is run time including any initialization phase; again copula factor models are excluded, this time due to frequent crashes.
The final method for comparison is the juxtaposition of the posterior predictive distribution with the original data. While the Bayesian Gaussian copula factor model does not have a full likelihood, we sample from an approximation to the posterior predictive distribution using the empirical marginal distributions; see Section 3.3 in \cite{murray2013bayescopula} for more details. 

For initialization, we use the default approach for all the methods. Specifically, Ellipsoid-Gaussian uses the aforementioned CTEF algorithm. In addition, as Ellipsoid-Gaussian uses a stochastic gradient Monte Carlo algorithm, a step size and a mini-batch size parameter need to be specified. We used a mini-batch size of 50 across different settings. For the step size, we used $10^{-4}$ for the data sets with $p = 3$; for higher dimensional data sets, we use $10^{-5}$ when the center is updated and $0.5 \times 10^{-5}$ when the center is fixed. While no tuning was necessary based on our experience, we notice that a smaller step size is preferable when the dimension is higher and/or when the center is fixed. In general we have observed that step sizes in $\{10^{-4}, 5 \times 10^{-5}, 10^{-5}\}$ tend to have good performance across a variety of data sets, and we recommend $10^{-5}$ as a default choice. 

\subsection{Simulated data sets\label{sec:simulation_setting}} 
For the simulation experiments, we generate four data sets, three from Ellipsoid-Gaussian distributions (two with curvature and one without), one from the Gaussian linear factor model and one from the hybrid Rosenbrock distribution \citep{Pagani21Rosenbrock}. The hybrid Rosenbrock was introduced as a benchmark to test the performance of Markov chain Monte Carlo on distributions with a curved and narrow shape. The density of a hybrid Rosenbrock random variable $x = (x_1, \ldots, x_{n_2, n_1})$ is proportional to
\begin{align}
    \exp\left\{-a_{ro} (x_1 - \nu)^2 - \sum_{j = 1}^{n_2} \sum_{i = 1}^{n_1} b_{ji}(x_{ji} - x_{j,i - 1}^2)^2\right\},\label{equ:rosenbrock}
\end{align}
up to normalizing constant, where $a_{ro}, b_{ji} \in \mathbb{R}^{+}.$ 
The settings of the simulated data sets are:
\begin{align*}
\text{very curved in $\mathbb{R}^{8}$:}&\quad{x}_i = \Lambda \eta_i + \epsilon_i, \quad \eta_i \sim \text{vMF}(\mu, 3), \quad \epsilon_i \sim \text{N}(0, 0.01 I_{8}), \quad k =4.\\ 
\text{approx. Gaussian in $\mathbb{R}^6$:}&\quad {x}_i = \Lambda \eta_i + \epsilon_i, \quad \eta_i \sim \text{vMF}(\mu, 30), \quad \epsilon_i \sim \text{N}(0, 0.4I_6),\quad k = 2.\\
\text{Gaussian linear factor in $\mathbb{R}^6$:}&\quad x_i =\Lambda \eta_i + \epsilon_i, \quad \eta_i \sim \text{N}(0, I_3), \quad \epsilon \sim \text{N}(0, 0.01 I_6),  \quad k = 3.\\
\text{hybrid Rosenbrock in $\mathbb{R}^3$:}&\quad a_{ro} = 0.2, \quad b = 0.05, \quad\nu = 1 \text{ in Equ.~\ref{equ:rosenbrock}}.
\end{align*}
The specification of $\Lambda$ is provided in the Supplementary Material \citep{song23eg_sm}.

When running experiments, we standardize each data set so that each variable has mean 0 and standard deviation 1, consistent with the practice of scaling variables, measured in different units, in applications. 
For very curved data in $\mathbb{R}^{8}$ generated from the Ellipsoid-Gaussian, we avoid standardization in order to assess performance in parameter estimation. 

\subsection{Simulation results}

\begin{figure}[ht]
\captionsetup[subfigure]{justification=Centering}
\begin{subfigure}[t]{0.24\textwidth}
	\includegraphics[width = \textwidth]{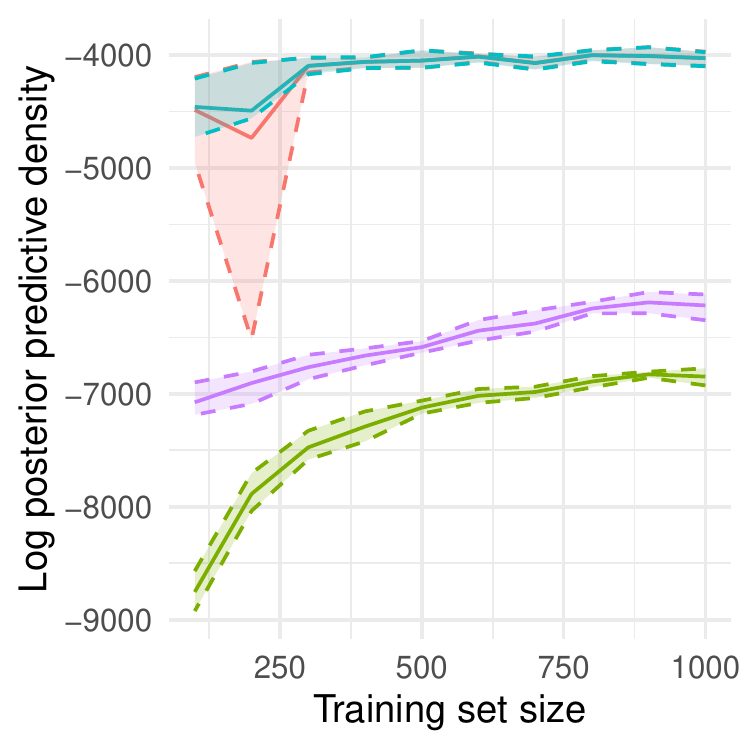}
	\caption{Very curved in $\mathbb{R}^{8}$\label{fig:8dk=4perf}}
\end{subfigure}
\begin{subfigure}[t]{0.24\textwidth}
	\includegraphics[width = \textwidth]{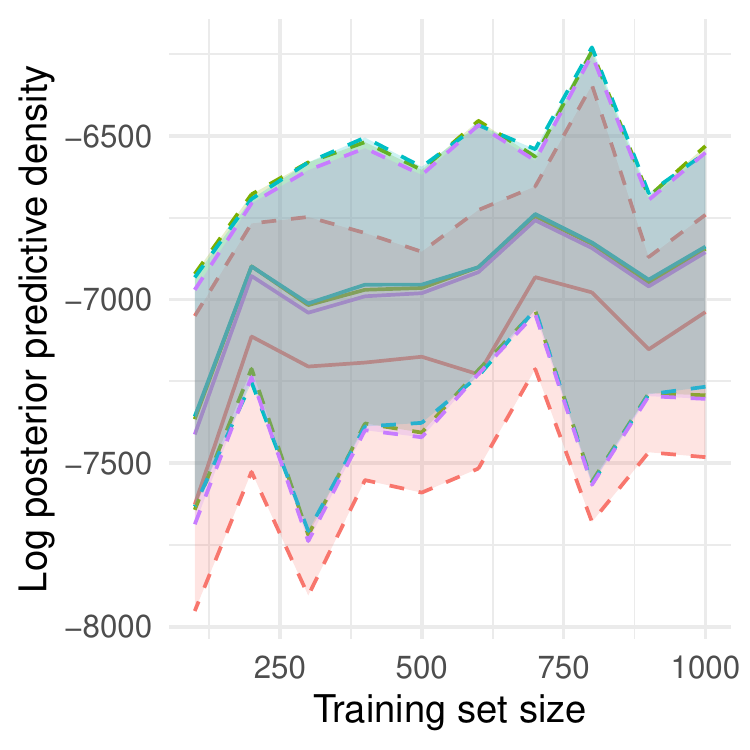}
	\caption{Approx. Gaussian in $\mathbb{R}^{6}$\label{fig:6degperf}}
\end{subfigure}
\begin{subfigure}[t]{0.24\textwidth}
\includegraphics[width = \textwidth]{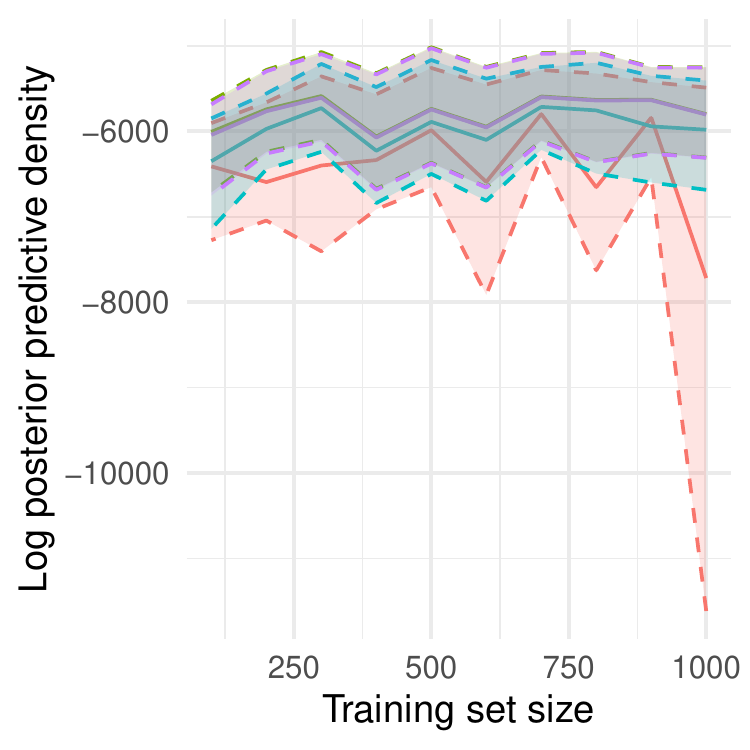}
\caption{Gaussian linear factor in $\mathbb{R}^6$\label{fig:glfperf}}
\end{subfigure}
\begin{subfigure}[t]{0.24\textwidth}
	\includegraphics[width = \textwidth]{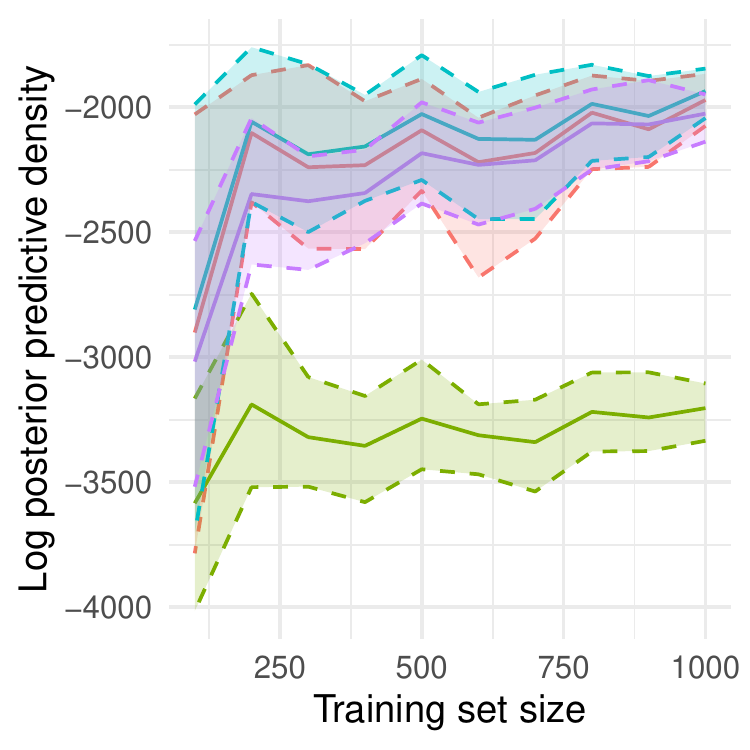}
	\caption{Hybrid Rosenbrock in $\mathbb{R}^3$\label{fig:3dhybridperf}}
\end{subfigure}
\caption{The log posterior predictive density of the test sets (1000 observations each) evaluated under each model as a function of the training set sizes with red, purple, blue and green corresponding to Ellipsoid-Gaussian (center fixed), Ellipsoid-Gaussian (center updated), Gaussian linear factor models and mixtures of factor analyzers, respectively. The dashed lines are 90\% and 10\% percentiles and the solid lines are the mean over 10 replicates.\label{fig:simulate_perf}}
\end{figure}

\begin{figure}[ht]
\centering
\begin{subfigure}[b]{0.24\textwidth}
\makebox[0pt][r]{\makebox[20pt]{\raisebox{42pt}{\rotatebox[origin=c]{90}{Hybrid Rosenbrock}}}}%
\includegraphics[width = \textwidth]{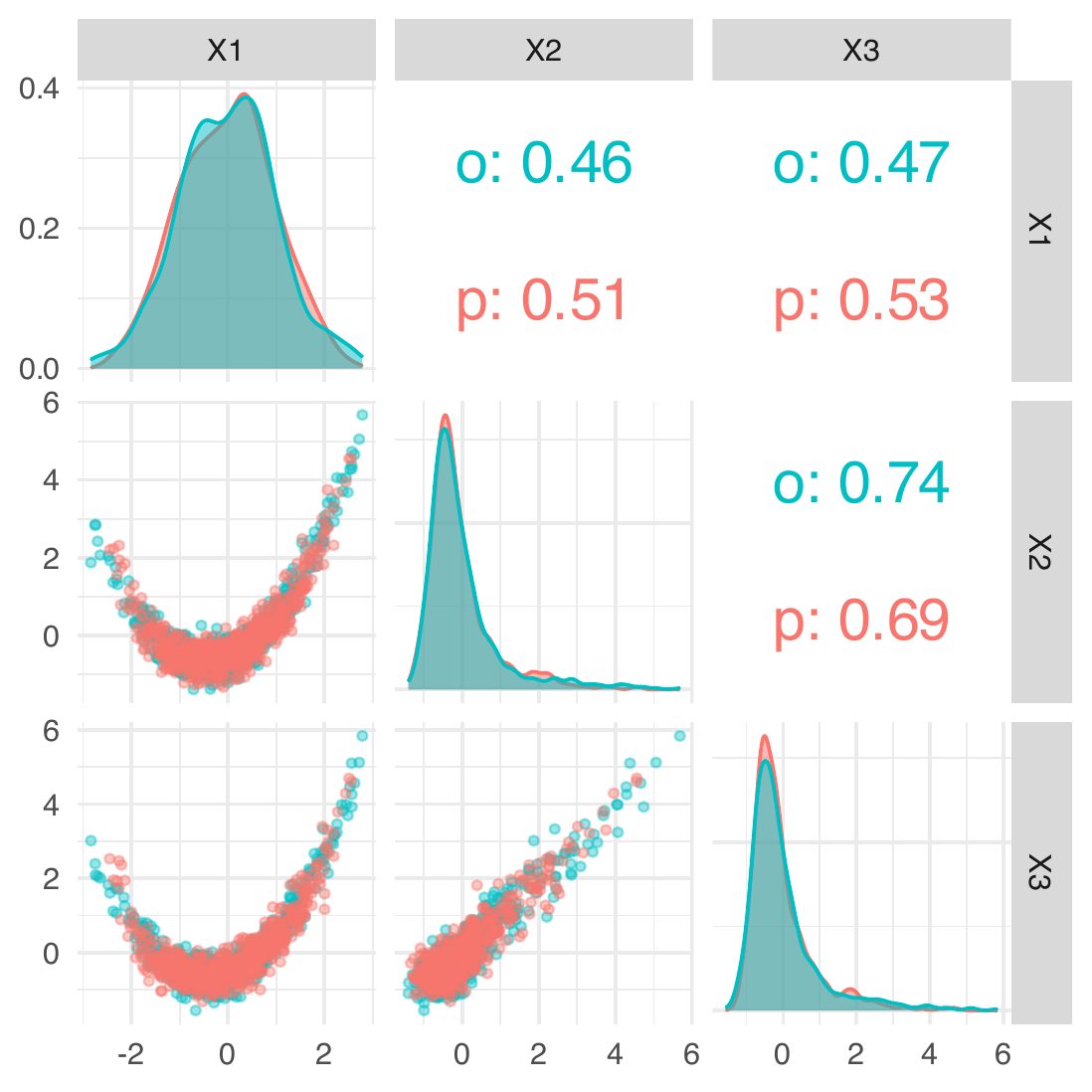}
\makebox[0 pt][r]{\makebox[20pt]{\raisebox{45pt}{\rotatebox[origin=c]{90}{Very curved}}}}%
  	\includegraphics[width = \textwidth]{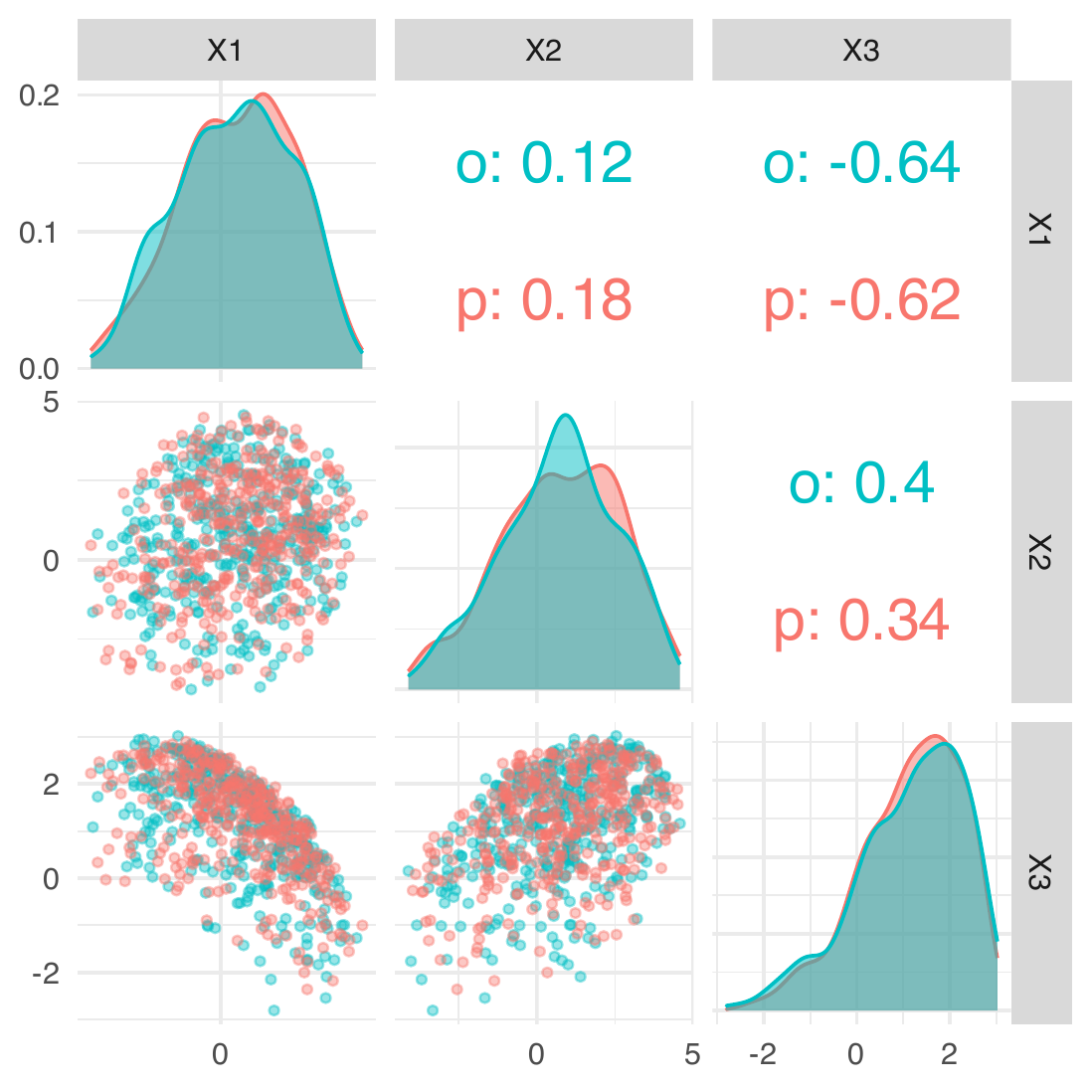}
	\caption{Ellipsoid-Gaussian (fixed center) \label{fig:eg_fix}}
	\end{subfigure}
\begin{subfigure}[b]{0.24\textwidth}
	\includegraphics[width = \textwidth]{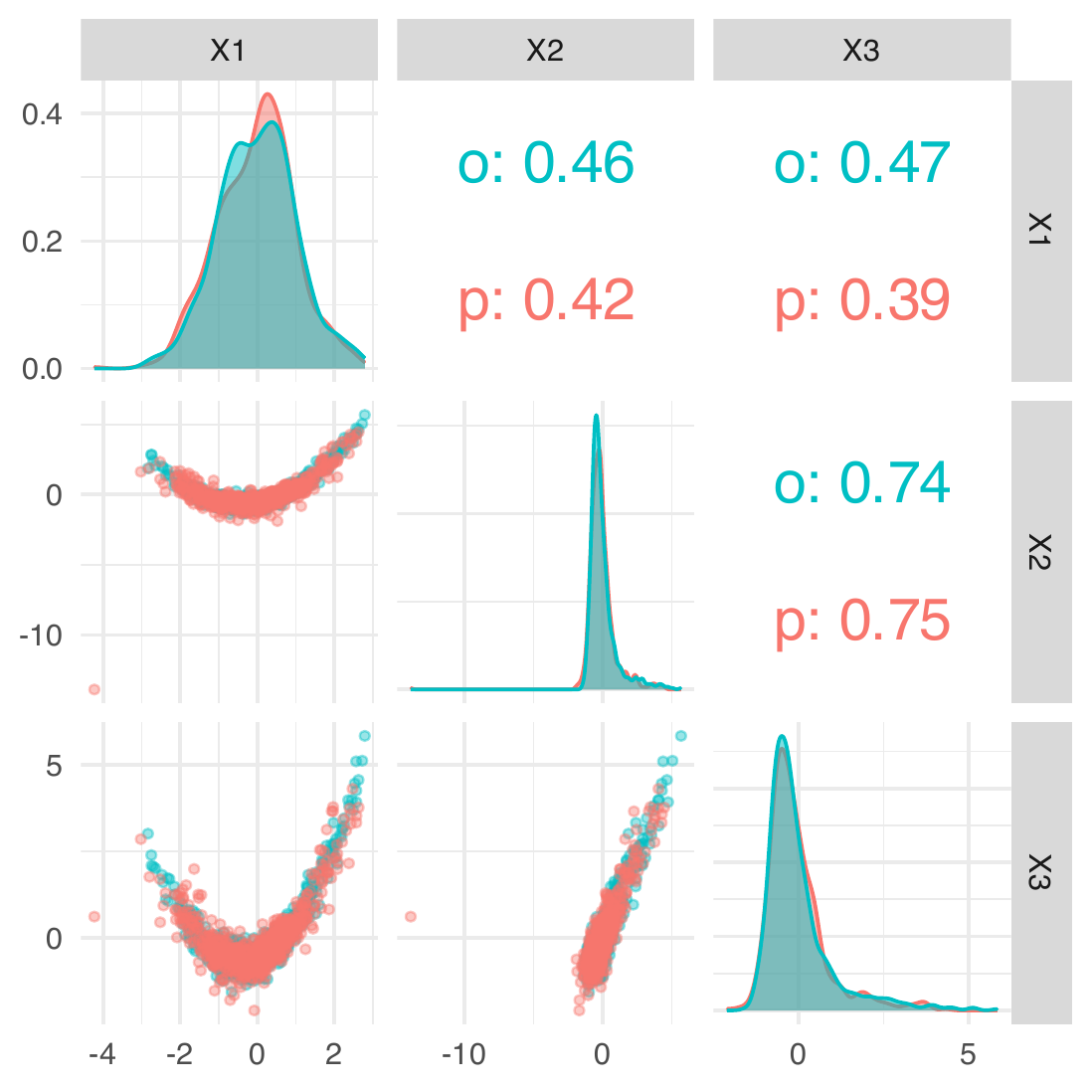}
 \includegraphics[width = \textwidth]{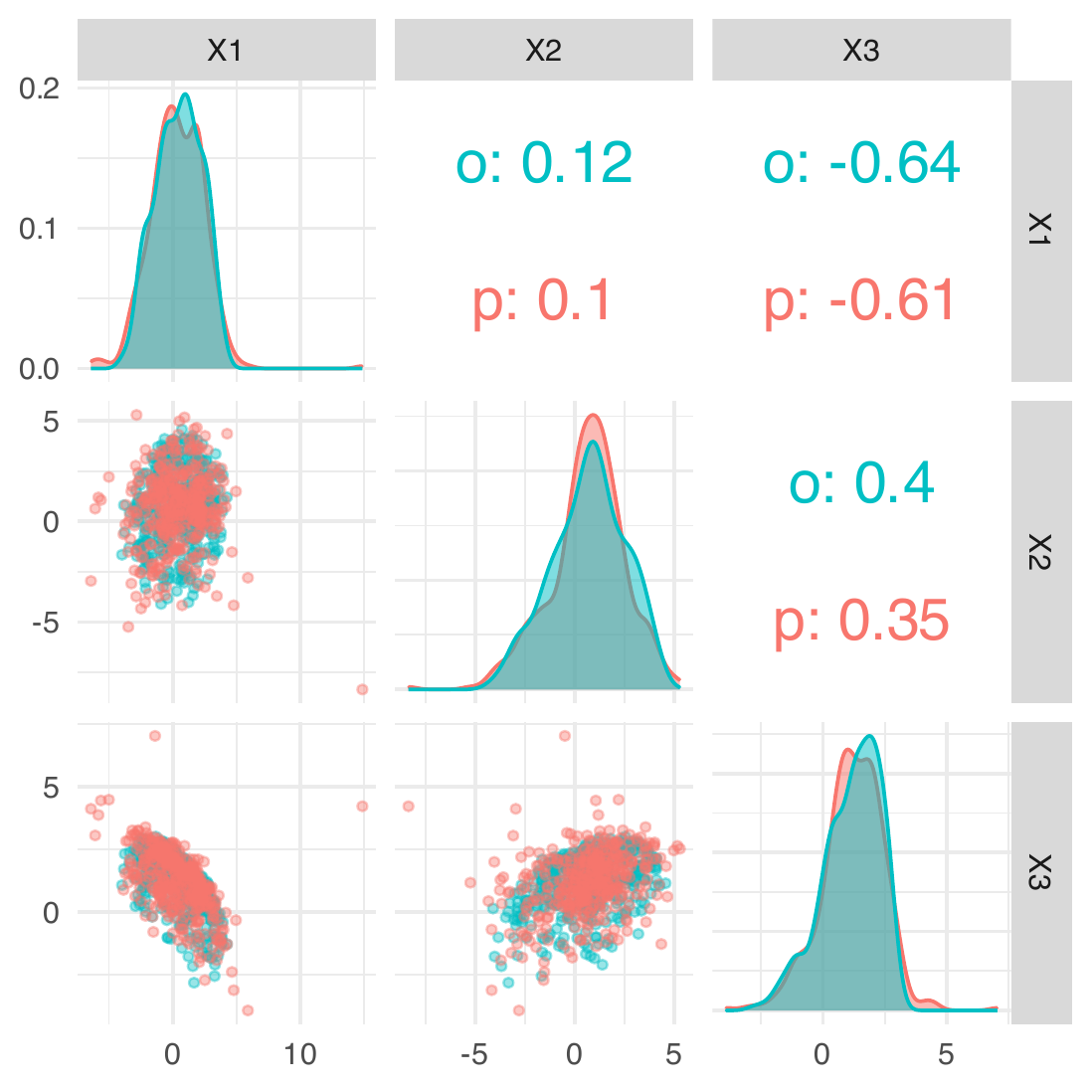}
	\caption{Mixtures of factor analyzers\label{fig:imifa}}
	\end{subfigure}
	\begin{subfigure}[b]{0.24\textwidth}
		\includegraphics[width = \textwidth]{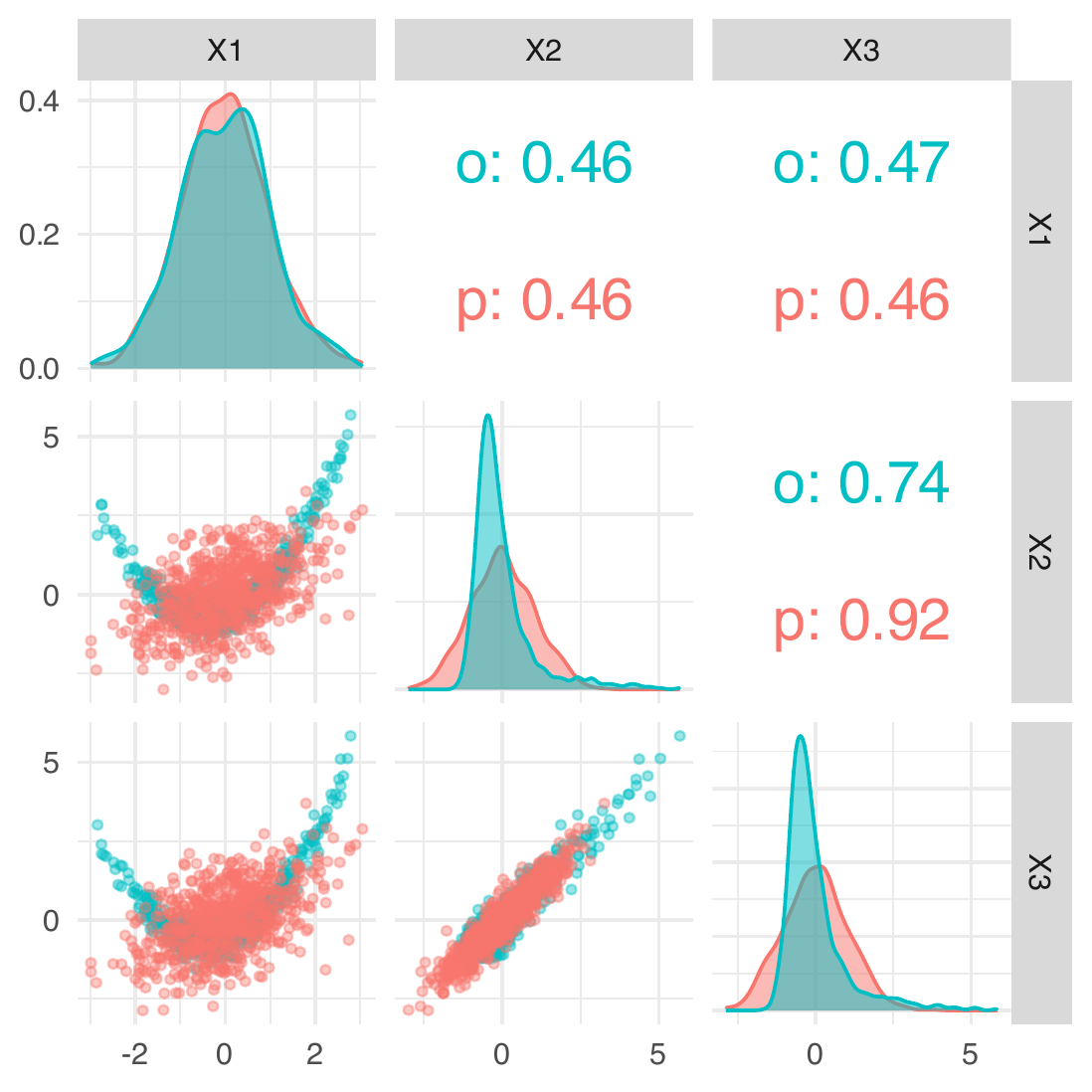}
  \includegraphics[width = \textwidth]{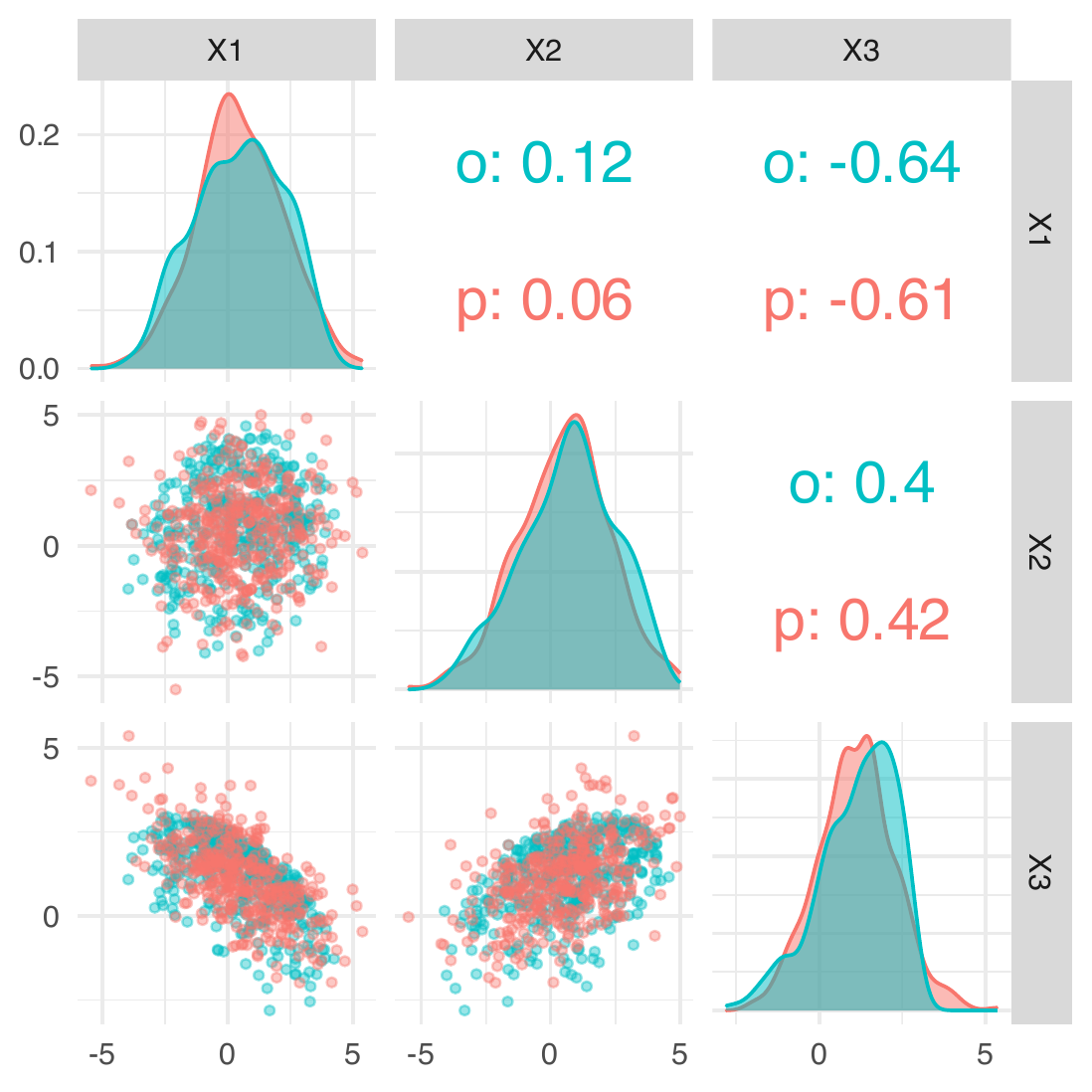}
	\caption{Gaussian linear factor models \label{fig:glf}}
	\end{subfigure}
 	\begin{subfigure}[b]{0.24\textwidth}
		\includegraphics[width = \textwidth]{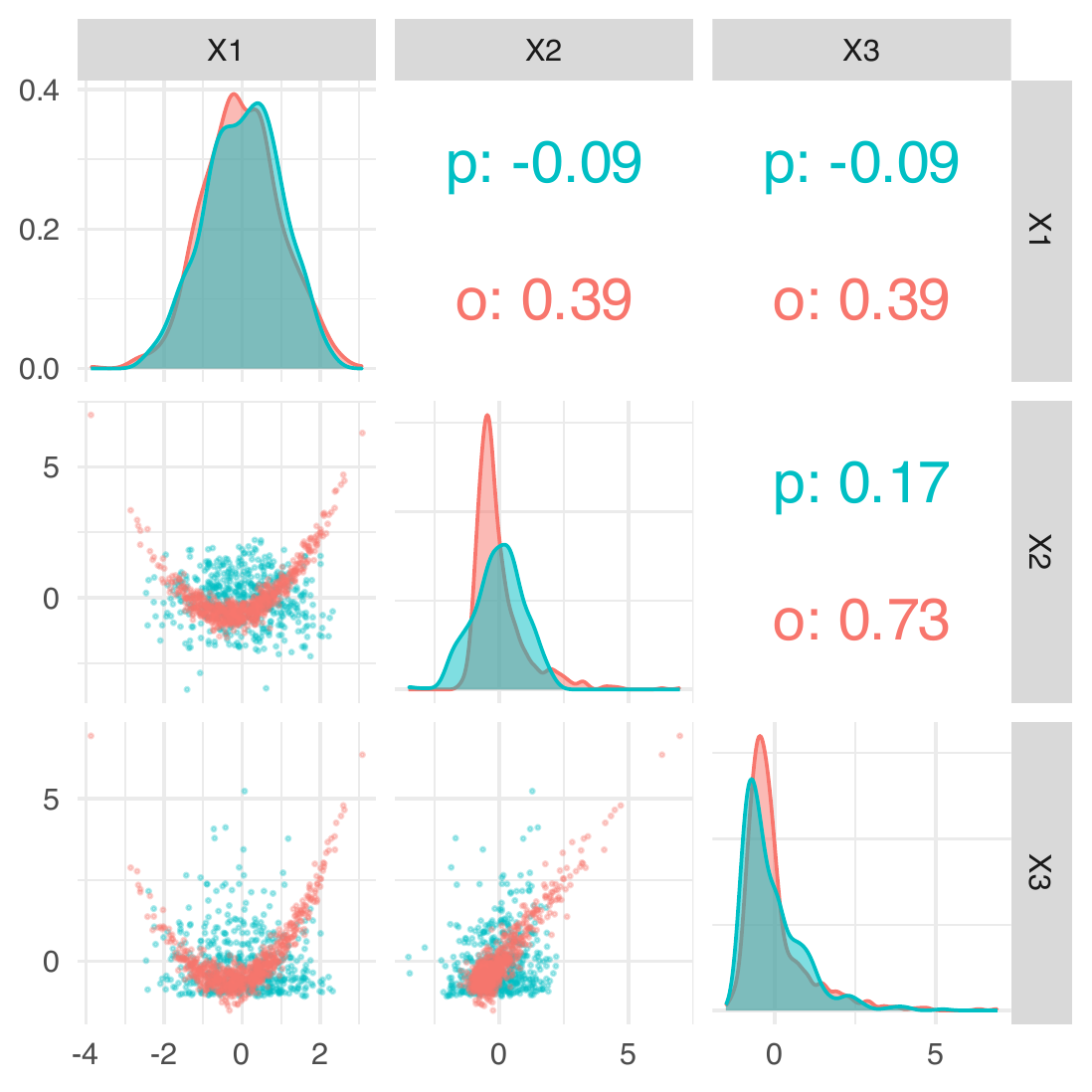}
  	\includegraphics[width = \textwidth]{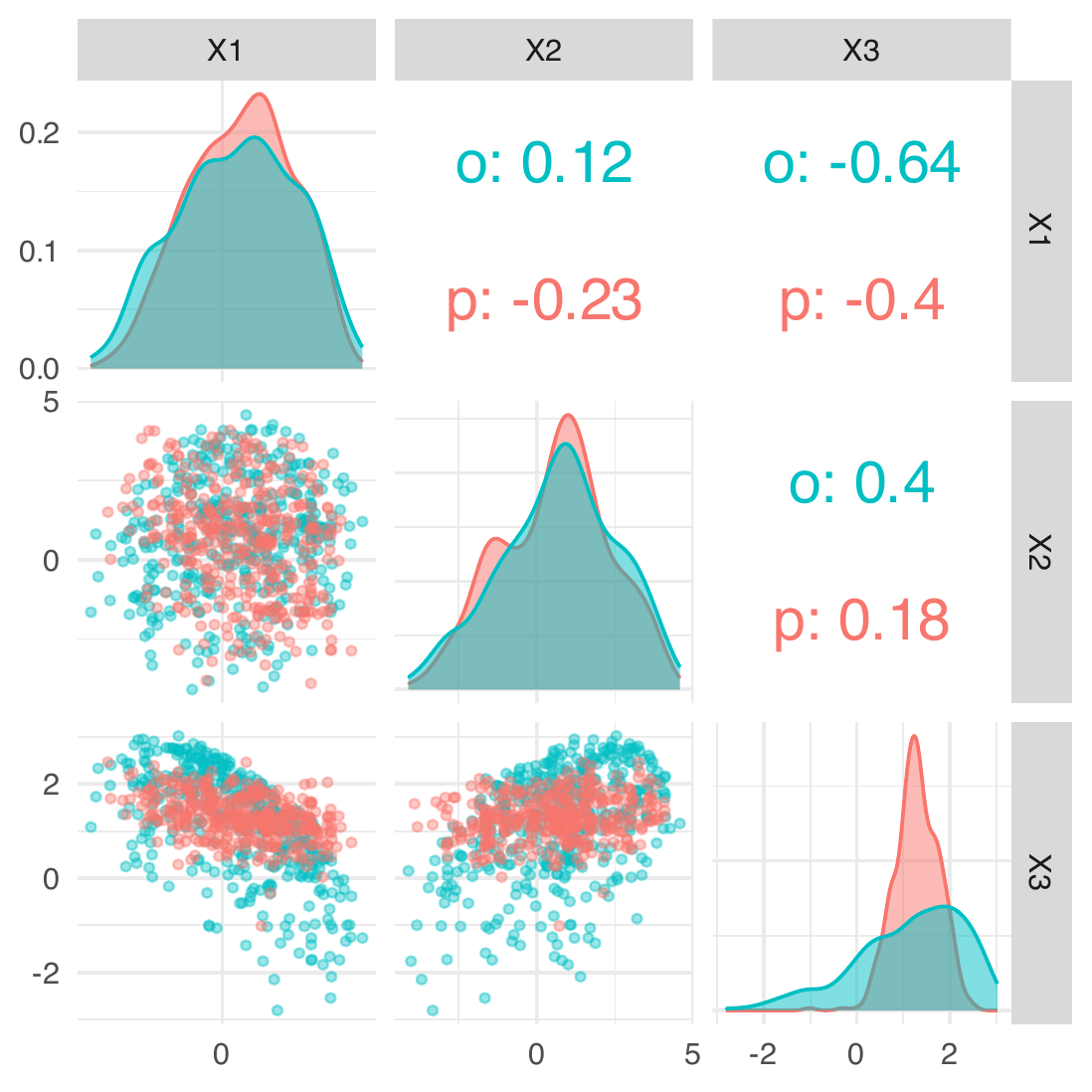}
	\caption{Gaussian Copula Factor Models \label{fig:bfa}}
	\end{subfigure}
\caption{The scatter plot matrices of the posterior predictive (green) juxtaposed with the original data (red) for each method, where the models were fitted to the hybrid Rosenbrock data sets in $\mathbb{R}^3$ with 700 observations (upper row) and  the very curved data sets in $\mathbb{R}^{8}$ with 400 observations (bottom row). The upper diagonals show correlations between the variables. Letters p and o stand for posterior predictive and original data respectively. \label{fig:simulate_perf_pairs}}
\end{figure}

\begin{figure}[ht]
\captionsetup[subfigure]{justification=Centering}
\begin{subfigure}[t]{0.24\textwidth}
	\includegraphics[width = \textwidth]{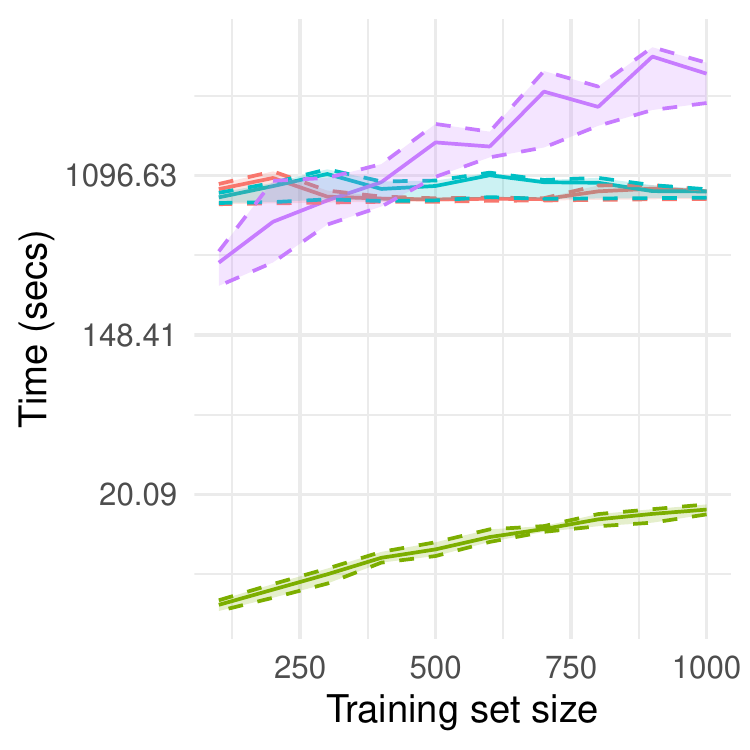}
	\caption{Very curved\label{fig:8dk=4time}}
\end{subfigure}
\begin{subfigure}[t]{0.24\textwidth}
\includegraphics[width = \textwidth]{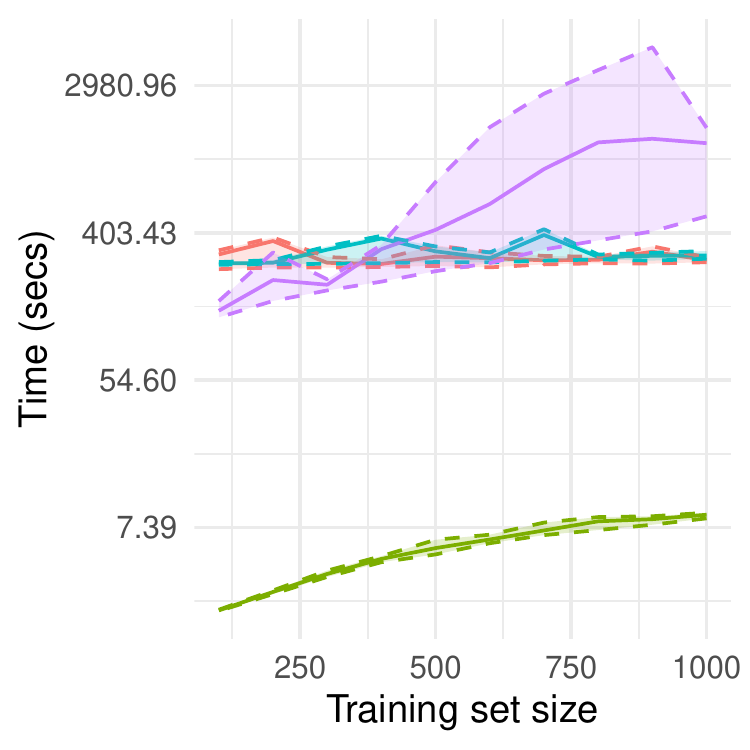}
\caption{Approx. Gaussian\label{fig:approxgaussiantime}}
\end{subfigure}
\begin{subfigure}[t]{0.24\textwidth}
\includegraphics[width = \textwidth]{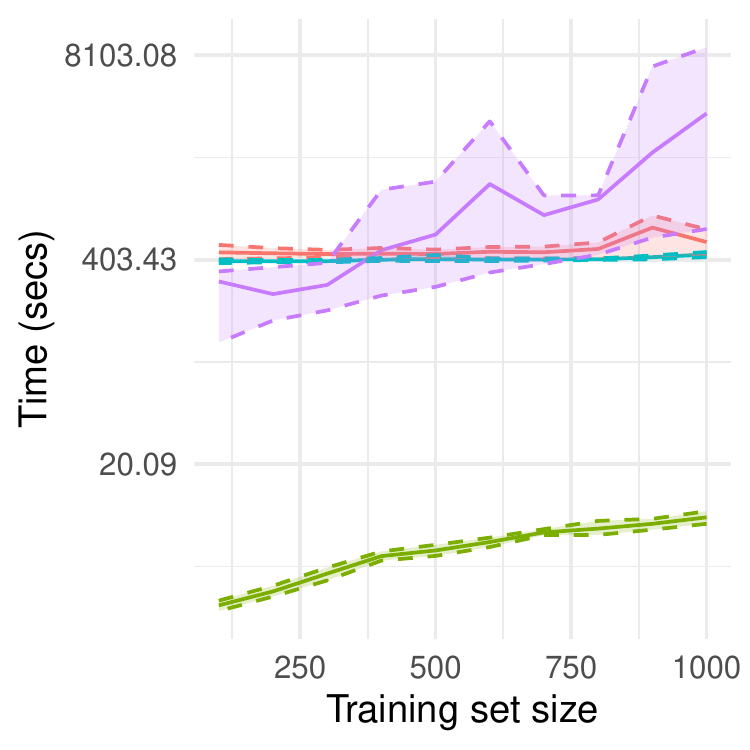}
\caption{Gaussian linear factor\label{fig:glftime}}
\end{subfigure}
\begin{subfigure}[t]{0.24\textwidth}
\includegraphics[width = \textwidth]{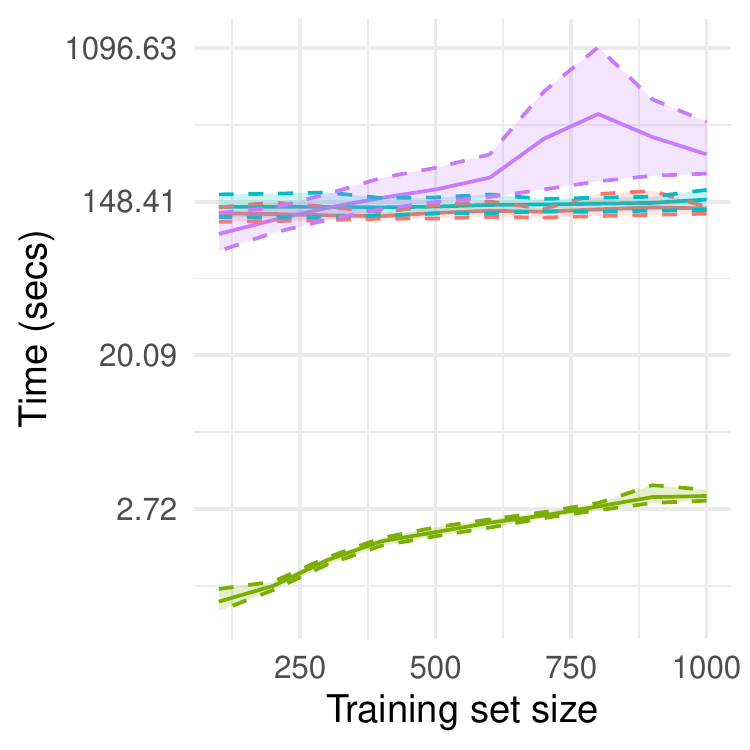}
\caption{Hybrid Rosenbrock\label{fig:rosenbrocktime}}
\end{subfigure}
\caption{The run time (including initialization) as a function of the training set sizes with red, purple, blue and green corresponding to Ellipsoid-Gaussian (center fixed), Ellipsoid-Gaussian (center updated), Gaussian linear factor models and mixtures of factor analyzers, respectively. The dashed lines are 90\% and 10\% percentiles and the solid lines are the mean over 10 replicates. \label{fig:simulate_time}}
\end{figure}

In the figures below, red, blue, green and purple represent Ellipsoid-Gaussian with center fixed, Ellipsoid-Gaussian with center updated, Gaussian linear factor models and 
mixtures of factor analyzers. Figure~\ref{fig:simulate_perf} and \ref{fig:simulate_time} show comparisons of out-of-sample log posterior predictive density and runtime, respectively. Figure~\ref{fig:simulate_perf_pairs} shows comparisons of posterior predictive distributions for two selected data sets, with results of the copula factor model included. More results (e.g. convergence diagnostics for Ellipsoid-Gaussian) can be found in the Supplementary Material \citep{song23eg_sm}.


For data sets with no curvature present, specifically approx.Gaussian in $\mathbb{R}^6$ and Gaussian linear factor in $\mathbb{R}^6$, Ellipsoid-Gaussian, Gaussian linear factor models and mixtures of factor analyzers perform comparably, as is shown in Figs.~\ref{fig:6degperf} and \ref{fig:glfperf}. This shows that it is safe to use Ellipsoid-Gaussian when it is not clear whether curvature is present in the data.

For data sets with curved dependence, Gaussian linear factor models and Gaussian copula factor models consistently fail to capture curvature as expected. While Gaussian linear factor models fail to capture curvature (see Fig.~\ref{fig:glf}), they perform well in terms of mean squared error. This highlights limitations of the mean squared error at capturing departures from normality and serves as a reminder that a model with low mean squared error can miss important features of the data. 

While mixtures of factor analyzers are able to capture curved dependence in the hybrid Rosenbrock example, its performance is only slightly better than a linear model for the very curved example, see Fig.~\ref{fig:8dk=4perf}. The method performed poorly despite the package selecting a large number of parameters. For example, for a data set of size 300, the package selected 3 mixture components each having factor loadings matrix of dimension 10 by 7, totaling 210 parameters just for the loadings matrices. We observed similar behavior for the hybrid Rosenbrock example; for example, for one data set of size 100, the package selected 10 mixture components, with each having factor loadings of 3 by 2, which resulted in 60 parameters just for the loadings matrices. This behavior suggests a lack of dimension reduction and interpretability.

In addition, the overuse of mixture components can result in spurious clusters or outliers, which is especially true when the model is fitted to small data; see the visualisation of the posterior predictive distributions in Figs.~\ref{fig:imifa}. It also leads to heavy computational costs, both in terms of run time and memory usage. Figure~\ref{fig:simulate_time} shows that mixtures of factor analyzers see dramatic increase in run time as the data size increases. The method also consumes much more memory as the data size increases---initially we attempted the very curved examples at $p = 10,$ but Rstudio experienced frequent crashes for large examples ($n \geq 900$). While Ellipsoid-Gaussian requires similar amount of run time for small data sets ($n \leq 300)$, the time does not increase with data sizes because of the use of a fixed mini-batch size.

\section{Applications\label{sec:application}}
We consider two applications, horse mussels and air quality data sets, where curved relationships are present (see scatter plot matrices in Fig.~\ref{fig:horsemussel} and \ref{fig:real_pairs_eg_fix}). We standardize both data sets prior to analysis. For all methods, we run 10,000 iterations of Markov chain Monte Carlo whenever possible and discard the first half as burn-in iterations. This was not possible for the copula factor model and mixtures of factor analyzers, when applying to the air quality data; details will be provided in Section \ref{sec:airquality}. 

For Ellipsoid-Gaussian, we use the same step sizes for both data sets, with $\epsilon =10^{-5}$ when the center is updated and $\epsilon = 0.5 \times 10^{-5}$ when the center is fixed. We process the posterior samples of factor loadings using the MatchAlign algorithm \citep{poworoznek2021align} to resolve rotational ambiguity and column label switching; we then visualize the posterior mean of the post-processed factor loadings. 

For comparison across models for the data, we visualize samples from posterior predictive distributions as with the simulation studies in Section~\ref{sec:simulation}. The supplementary material \citep{song23eg_sm} contains additional results for Ellipsoid-Gaussian, including diagnostics plots for posterior sampling and a study of the relationships between two variables of interest while holding the remaining ones at their sample means.

\subsection{Horse mussel data} 
The horse mussels data can be found in our R package \texttt{ellipsoidgaussian} \citep{song23eg}, and contains measurements of 201 horse mussels taken from five sites in New Zealand in December 1984 \citep{CamdenMike1989Tdb, cook1998regression}. The measurements include shell width $W$, length $L$ and height $H$ (each in millimeters), and shell mass $S$ and muscle mass $M$ (both in grams). An interest of the study lies in studying the relationship of muscle mass, the edible portion of a horse mussel, with other variables. For Ellipsoid-Gaussian, we use $k = 3$ factors to model the data.

\begin{figure}
\centering
\makebox[0pt][r]{\makebox[20pt]{\raisebox{60pt}{\rotatebox[origin=c]{90}{Horse mussel data}}}}%
\includegraphics[width = 0.28\textwidth]{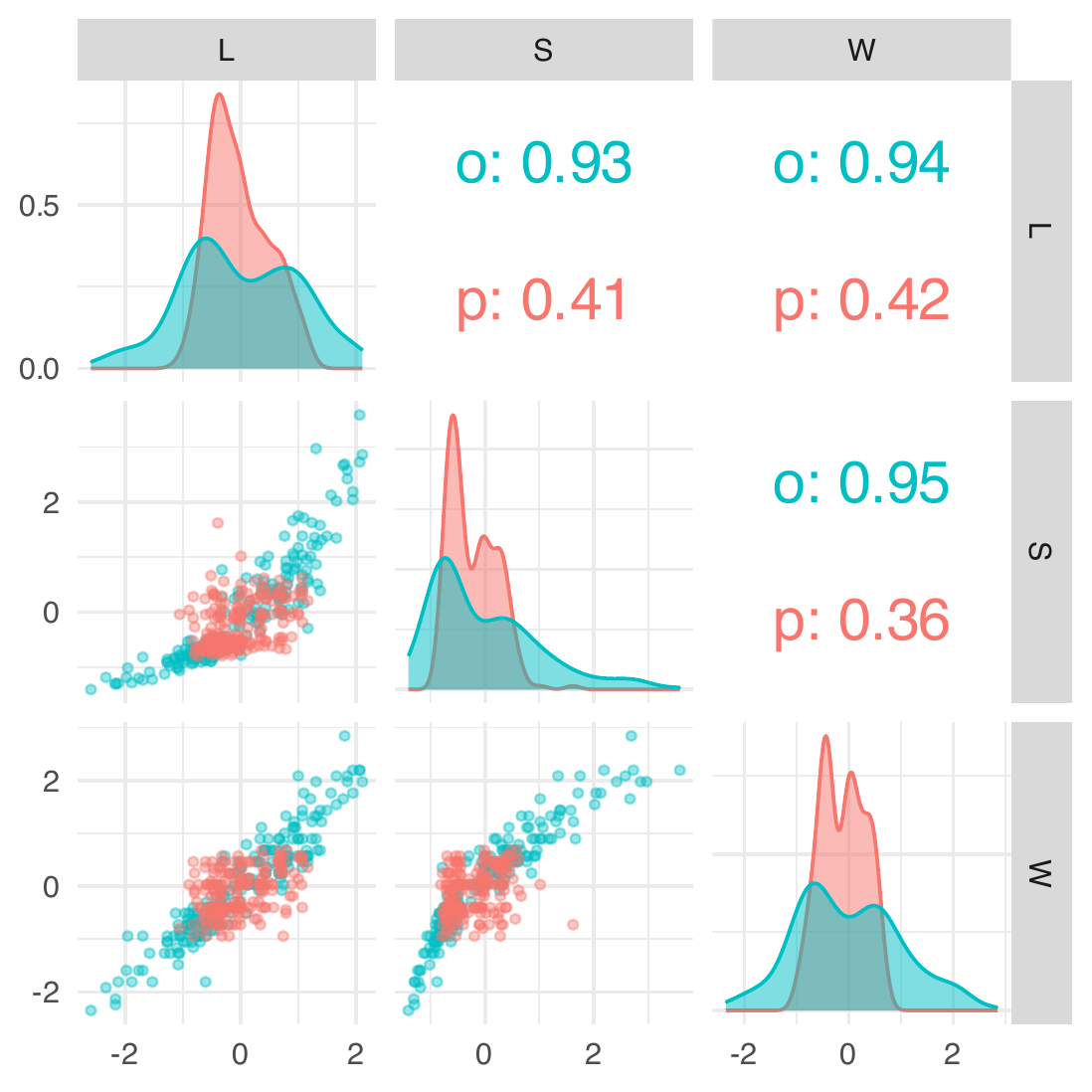}
\caption{Bayesian copula factor model result: the scatter plot matrices of the posterior predictive (green) juxtaposed with the original data (red). The upper diagonals are correlations between the variables. Letters p and o stand for posterior predictive and original data respectively.\label{fig:horse_mussel_bfa}}
\end{figure}
\begin{figure}[!htb]
\captionsetup[subfigure]{justification=Centering}
\begin{subfigure}[t]{0.24\textwidth}
\makebox[0pt][r]{\makebox[30pt]{\raisebox{45pt}{\rotatebox[origin=c]{90}{Horse mussel data}}}}%
	\includegraphics[width=\textwidth]{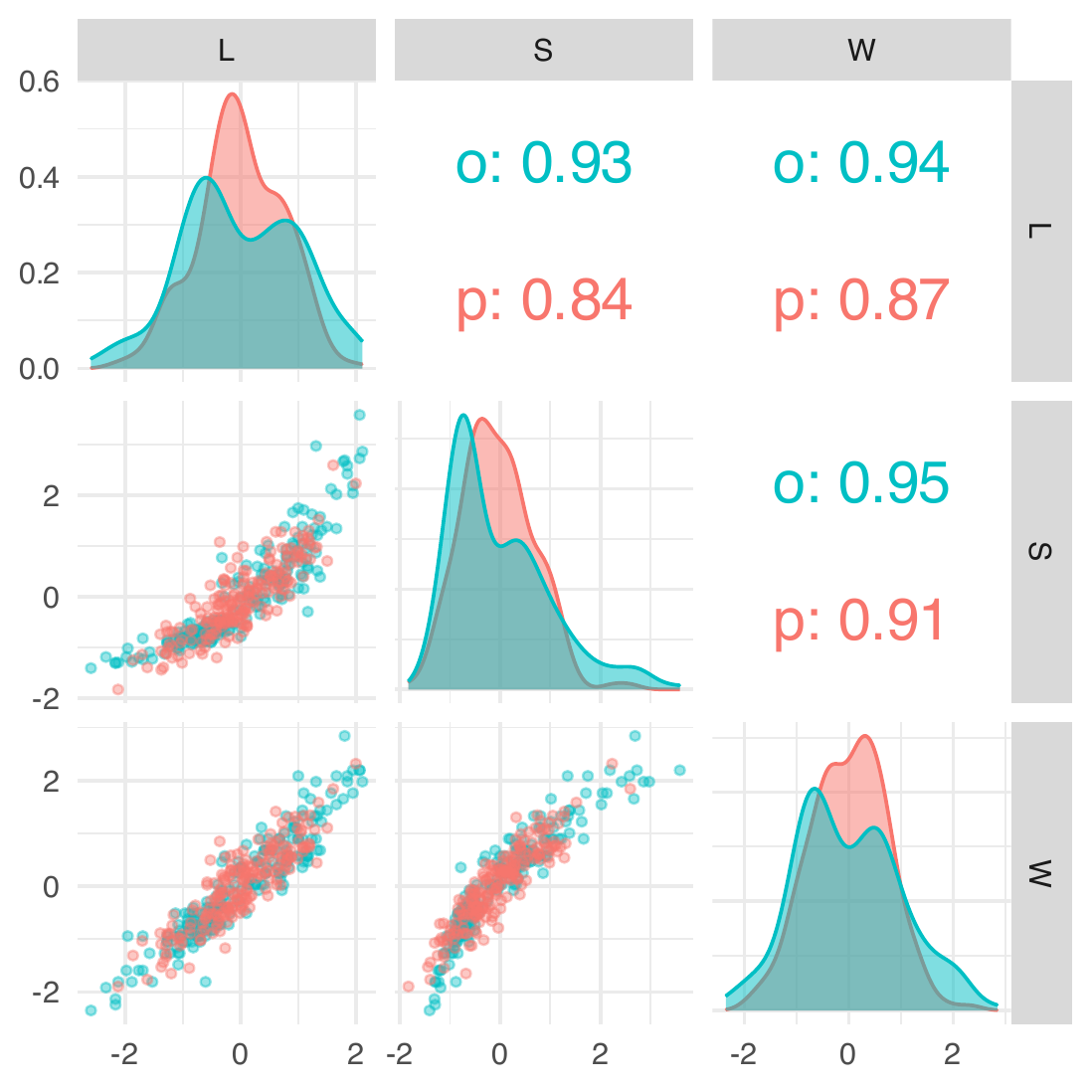}
	\makebox[0pt][r]{\makebox[30pt]{\raisebox{45pt}{\rotatebox[origin=c]{90}{Air quality data}}}}%
	\includegraphics[width=\textwidth]{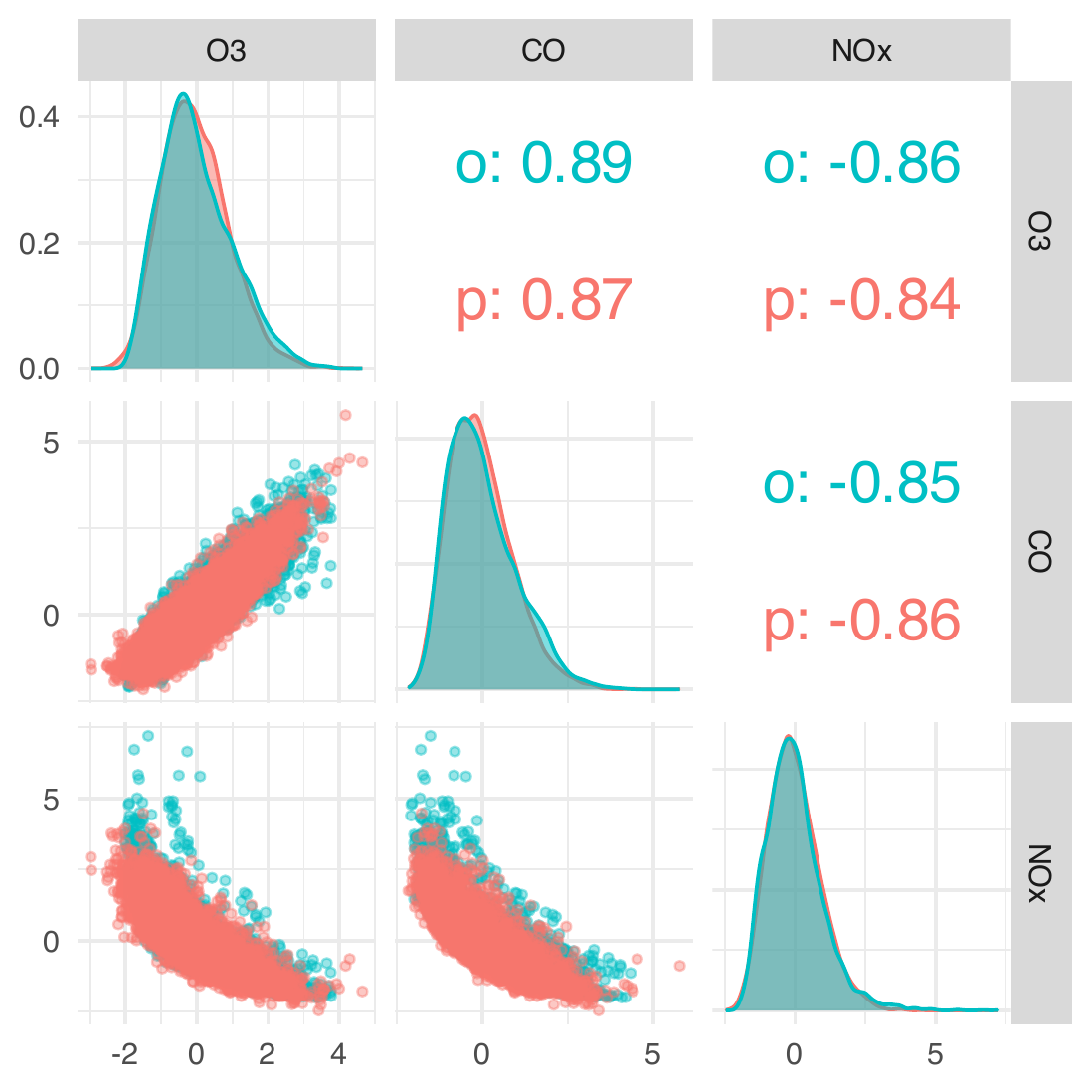}
	\caption{Ellipsoid-Gaussian (fix center) \label{fig:real_pairs_eg_fix}}
\end{subfigure}
\begin{subfigure}[t]{0.24\textwidth}
	\includegraphics[width=\textwidth]{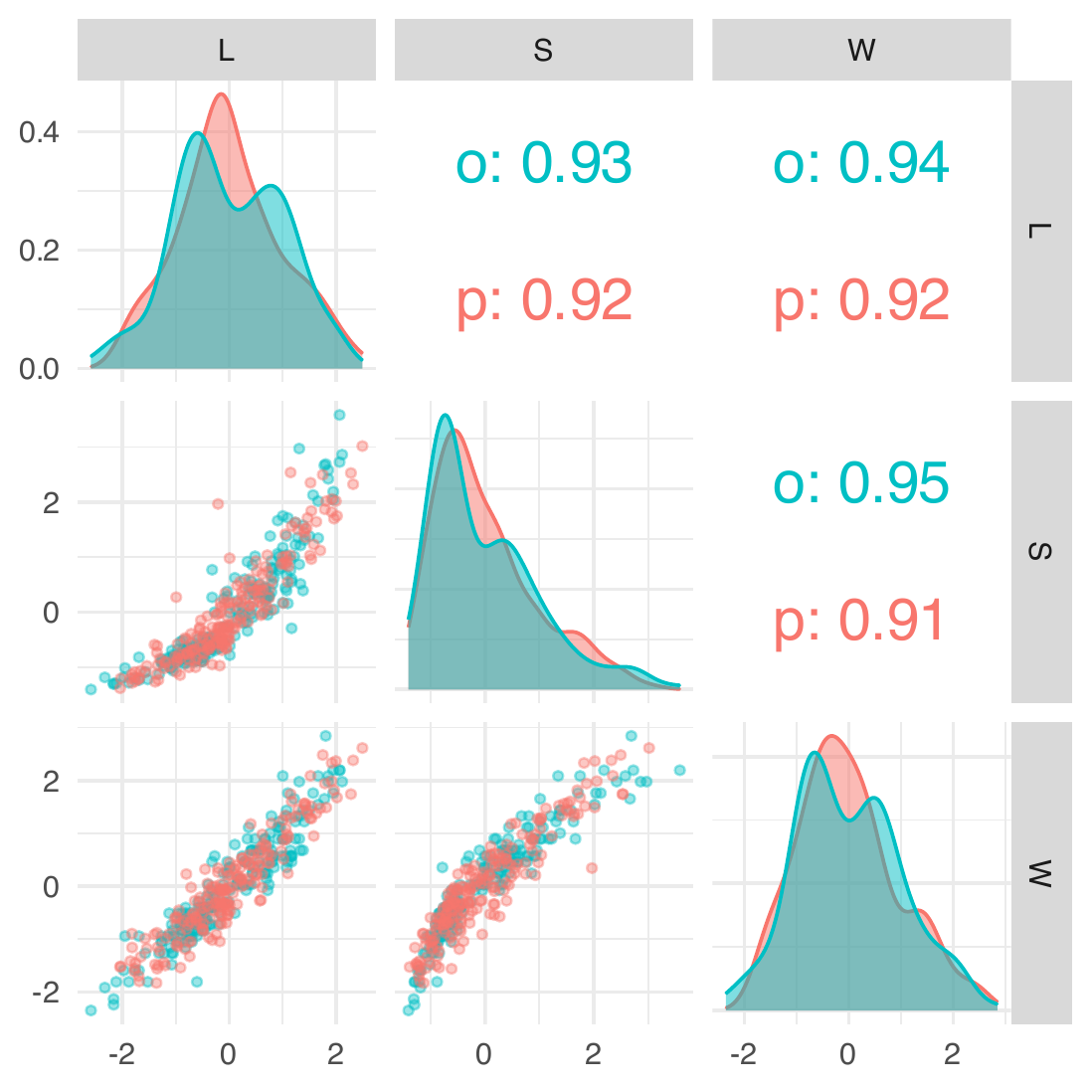}
	\includegraphics[width=\textwidth]{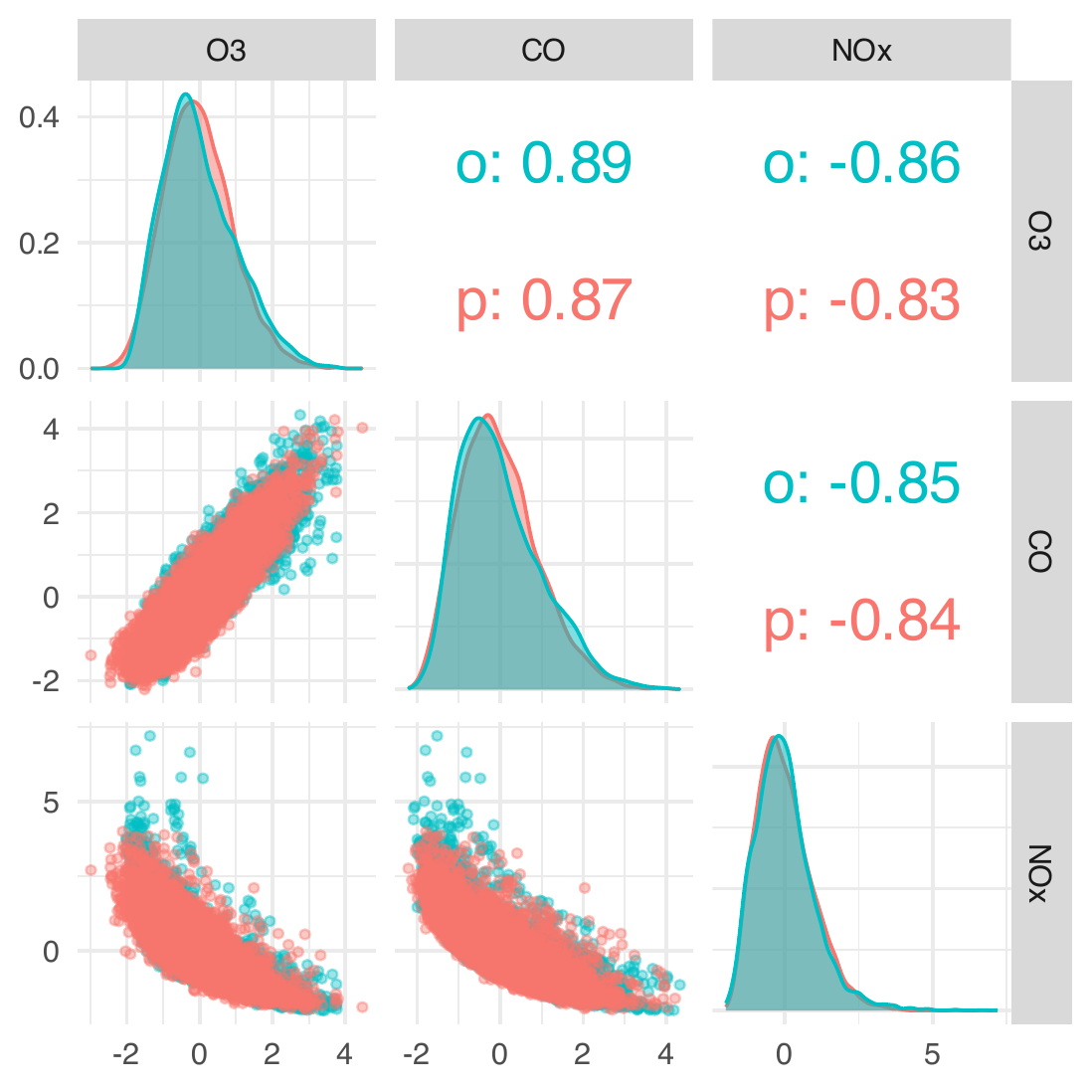}
	\caption{Ellipsoid-Gaussian (update center) \label{fig:real_pairs_eg_update}}
\end{subfigure}
\begin{subfigure}[t]{0.24\textwidth}
	\includegraphics[width=\textwidth]{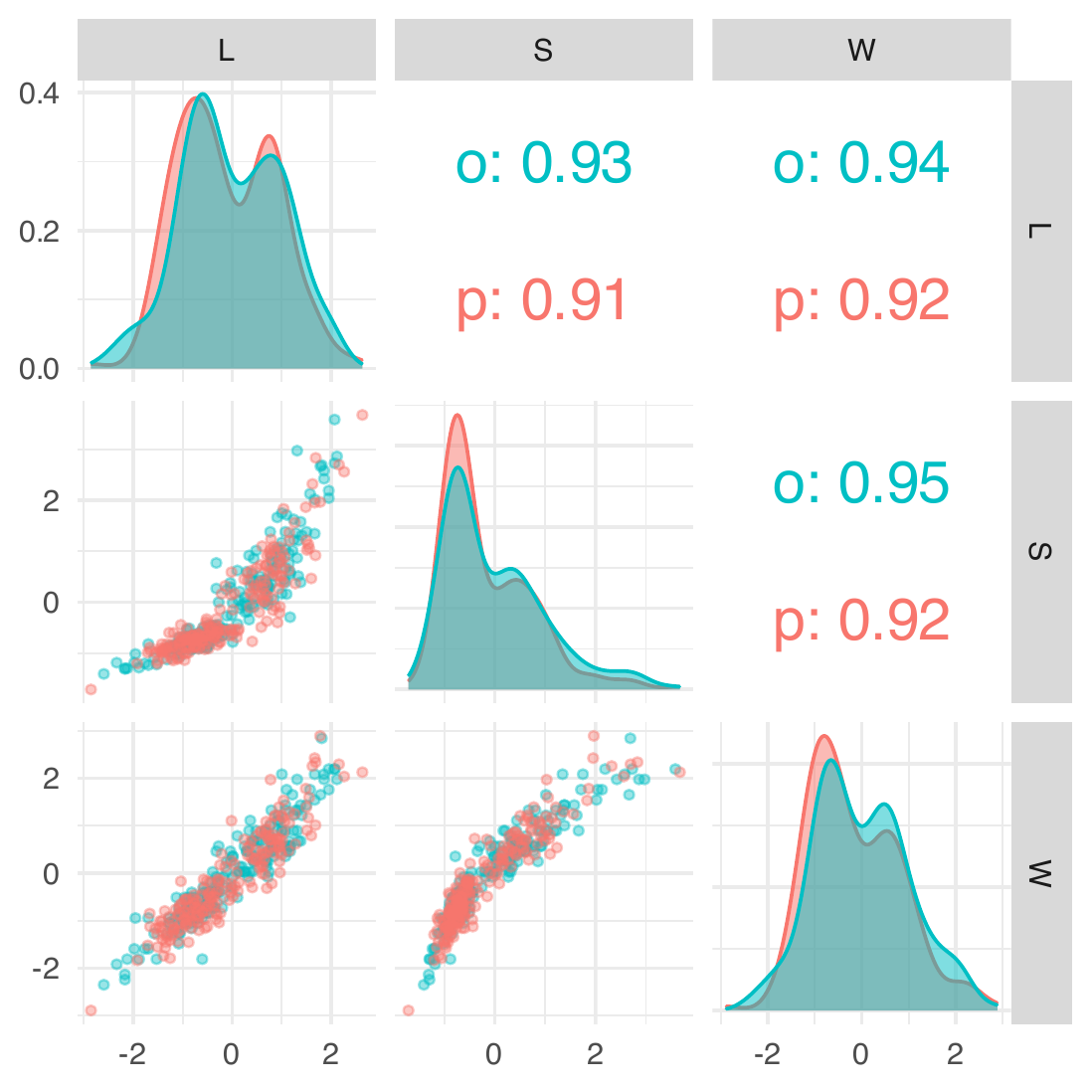}
	\includegraphics[width=\textwidth]{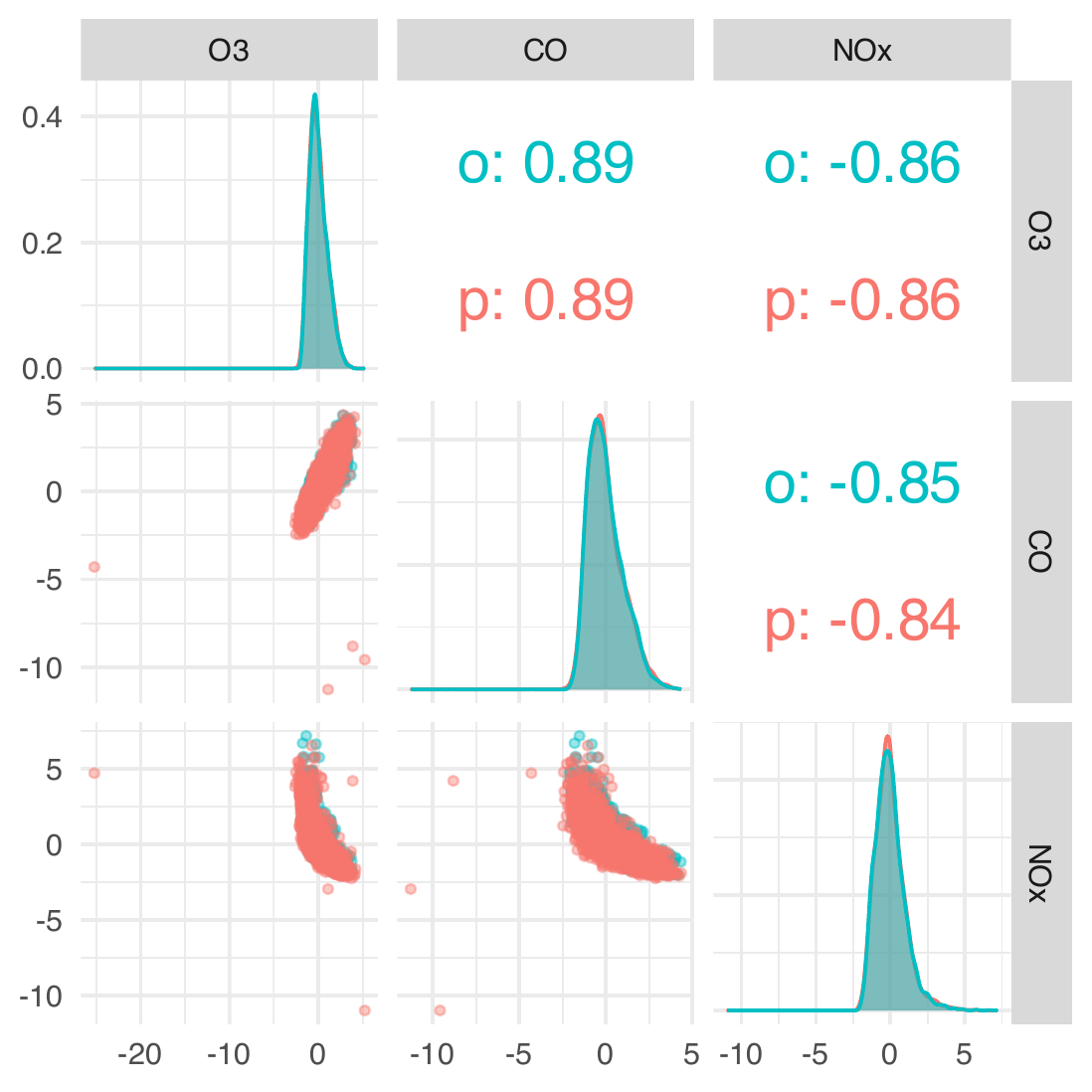}
	\caption{Mixtures of factor analyzers\label{fig:real_pairs_imifa}}
\end{subfigure}
\begin{subfigure}[t]{0.24\textwidth}
	\includegraphics[width=\textwidth]{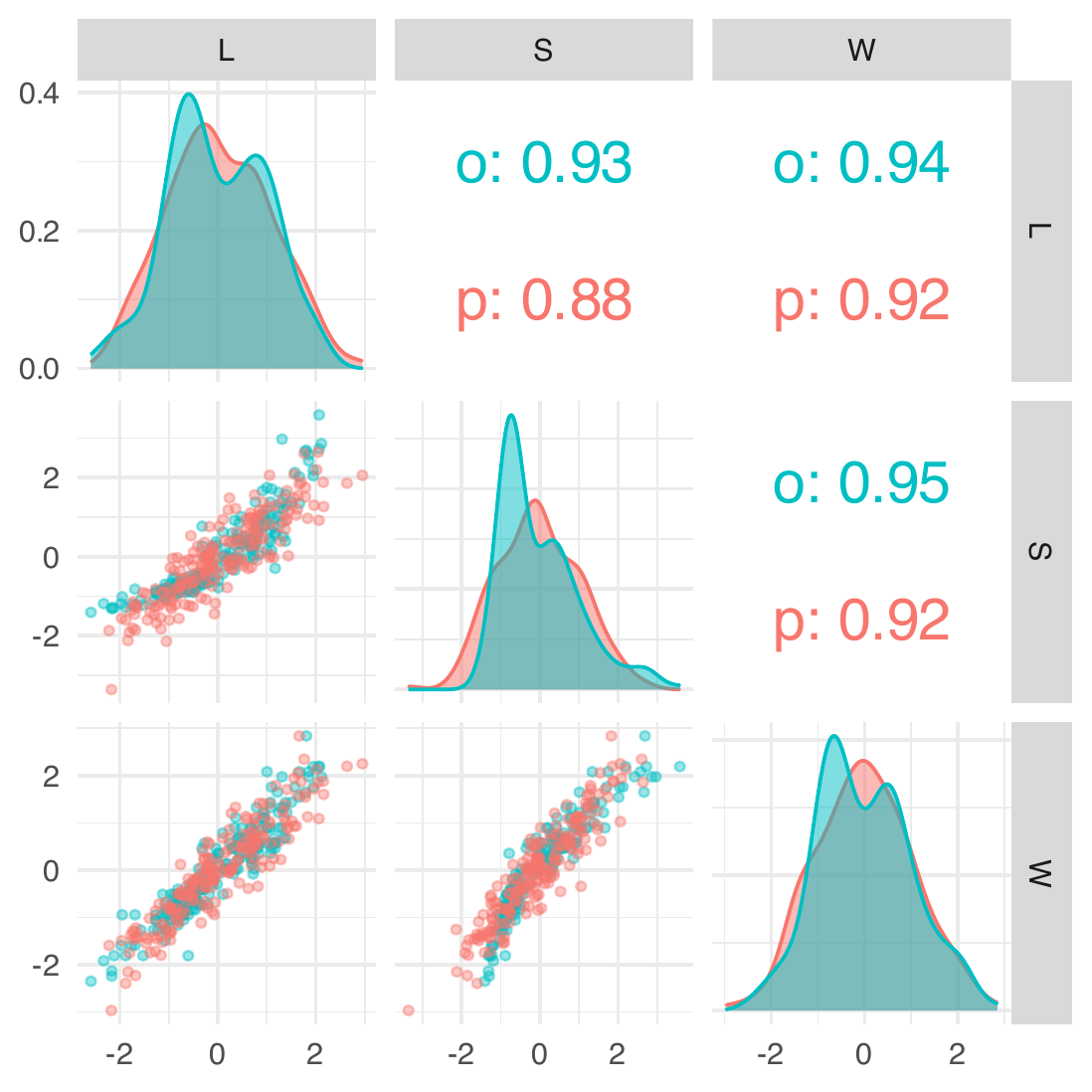}
	\includegraphics[width=\textwidth]{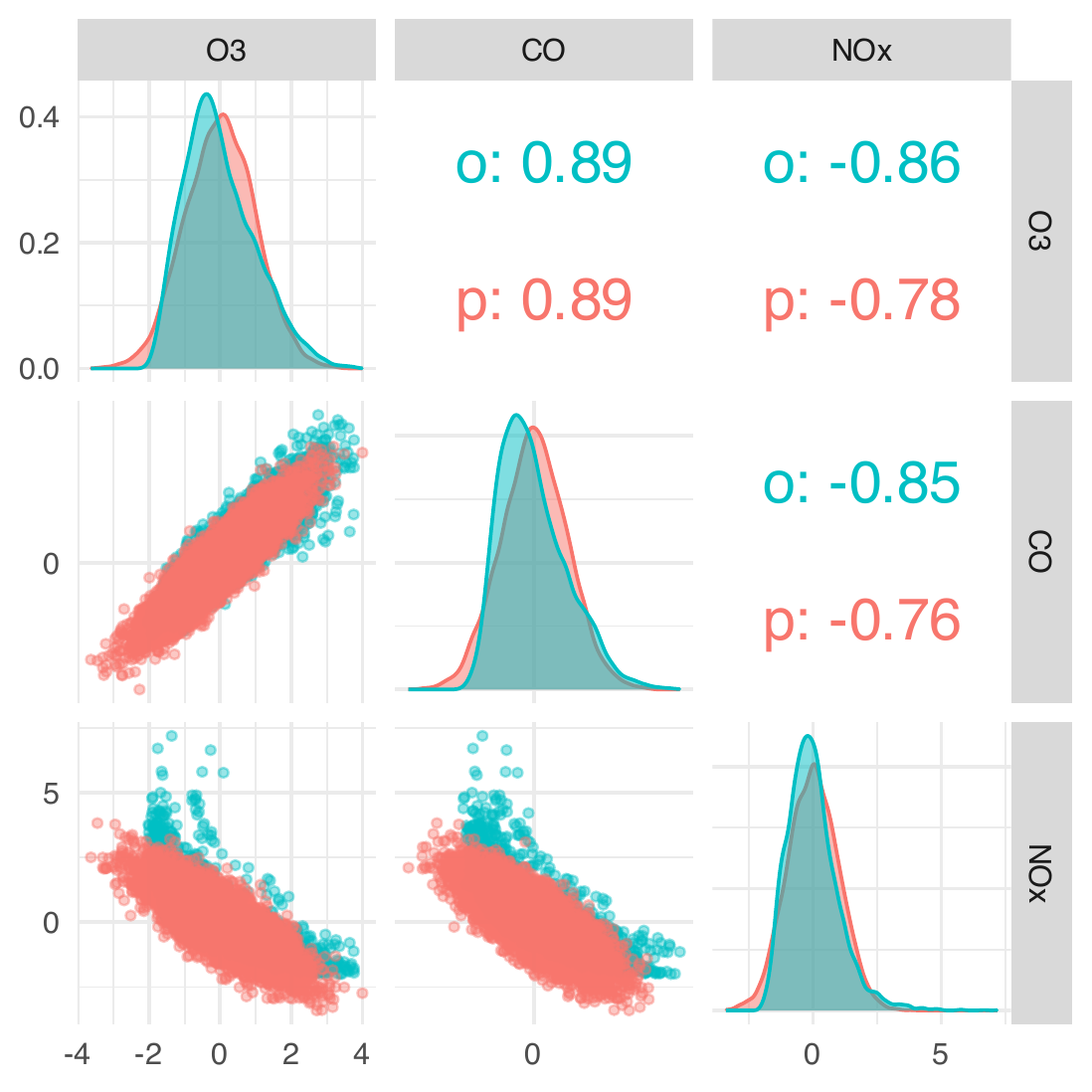}
	\caption{Gaussian linear factor model\label{fig:real_pairs_glf}}
\end{subfigure}
\caption{The scatter plot matrices of the posterior predictive (green) juxtaposed with the original data (red). The upper diagonals are correlations between the variables. Letters p and o stand for posterior predictive and original data respectively.\label{fig:real_pairs}}
\end{figure}
\begin{figure}[!htb]
\centering
\captionsetup[subfigure]{justification=Centering}
\begin{subfigure}[t]{0.67\textwidth}
\makebox[0pt][r]{\makebox[20pt]{\raisebox{60pt}{\rotatebox[origin=c]{90}{Horse mussel data}}}}%
\includegraphics[width = 0.48\textwidth]{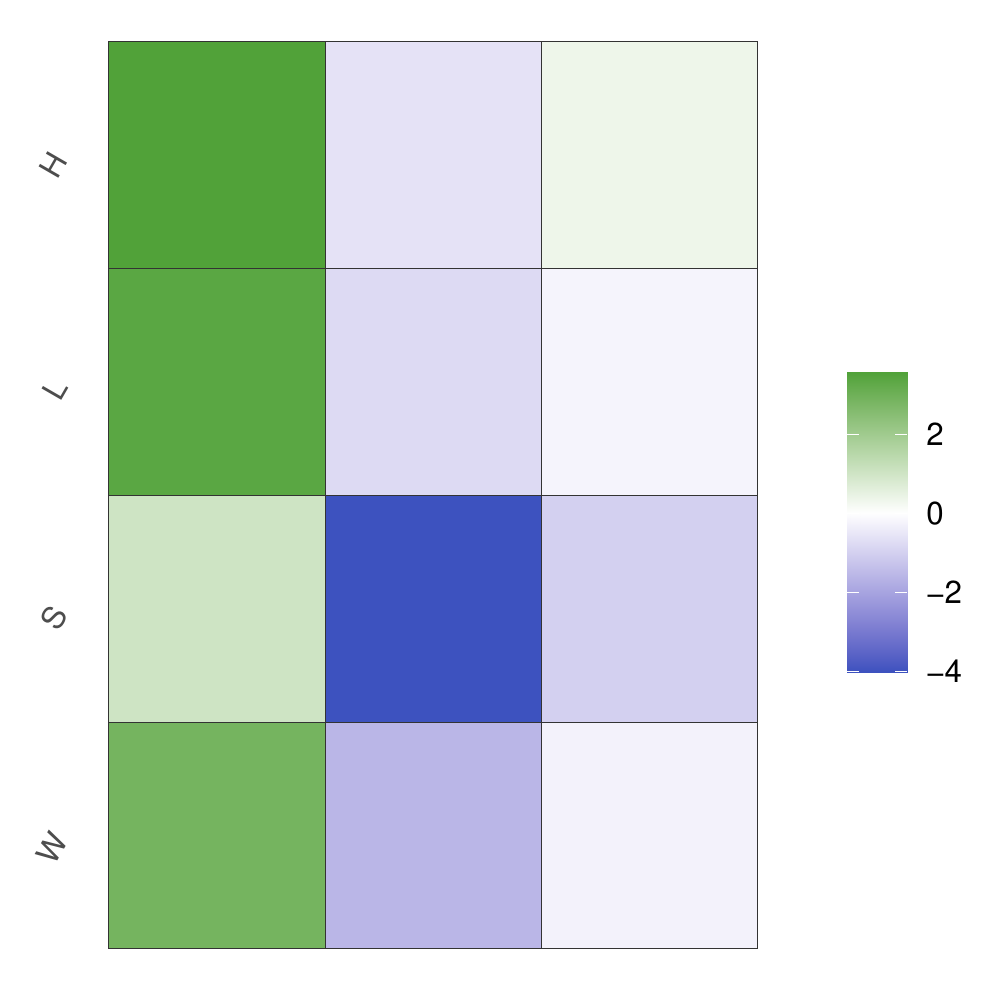}
\includegraphics[width = 0.48\textwidth]{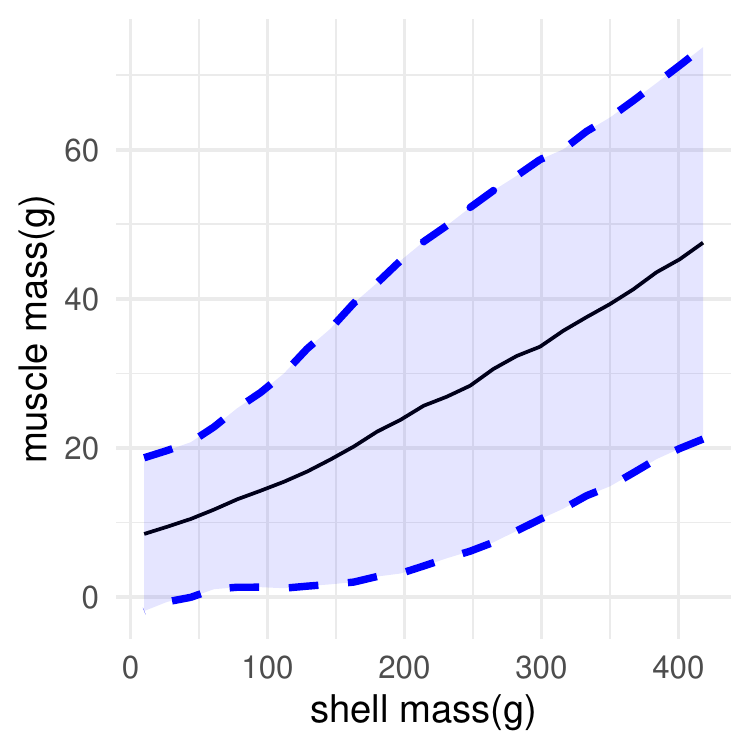}
\makebox[0pt][r]{\makebox[20pt]{\raisebox{60pt}{\rotatebox[origin=c]{90}{Air quality data}}}}%
\includegraphics[width = 0.48\textwidth]{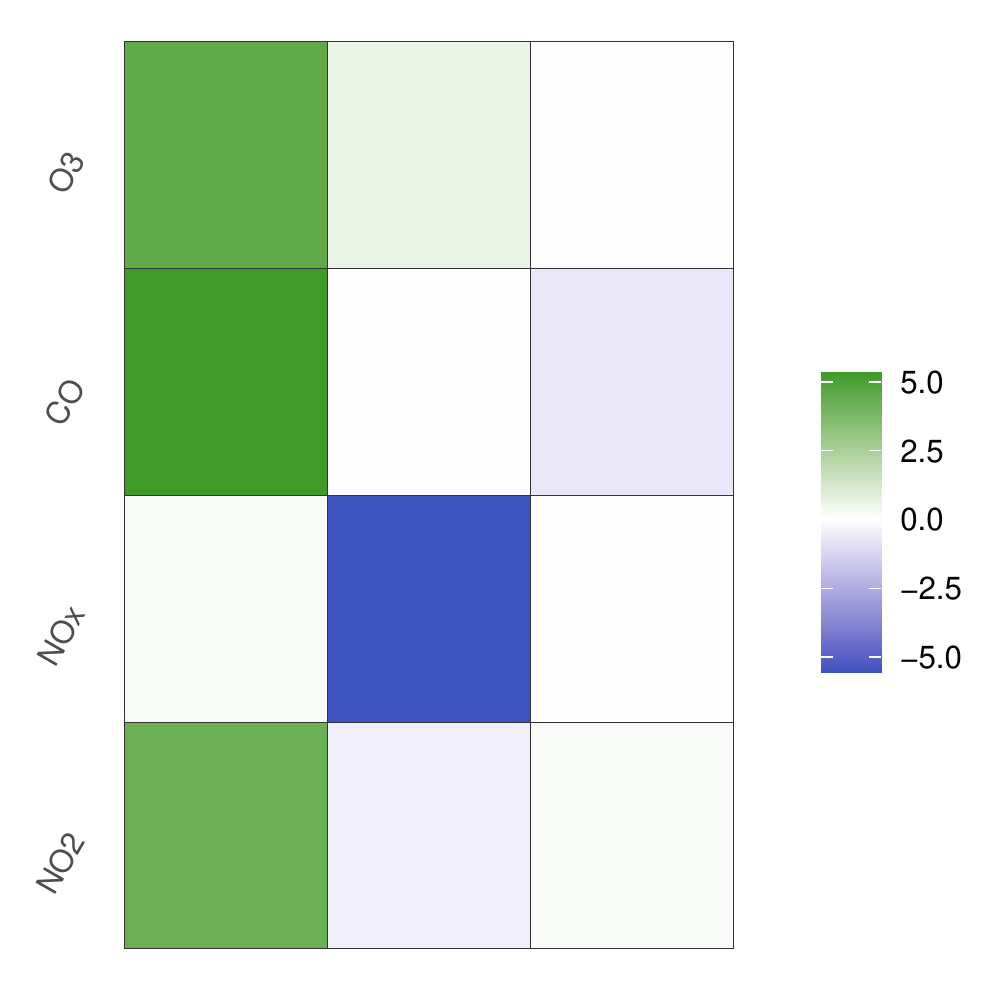}
\includegraphics[width = 0.48 \textwidth]{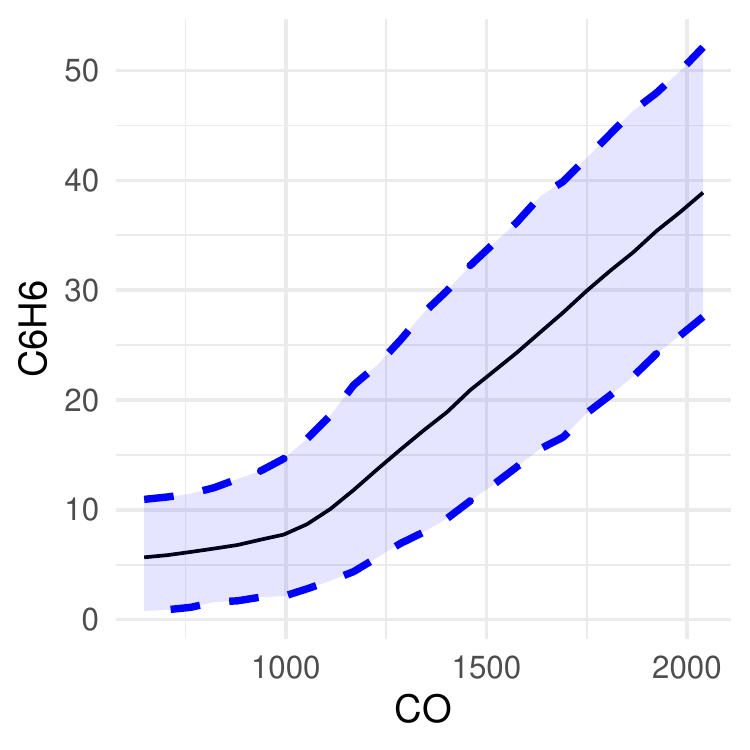}
\caption{Update center \label{fig:loadings_update}}
\end{subfigure}
\begin{subfigure}[t]{0.32\textwidth}
\includegraphics[width = \textwidth]{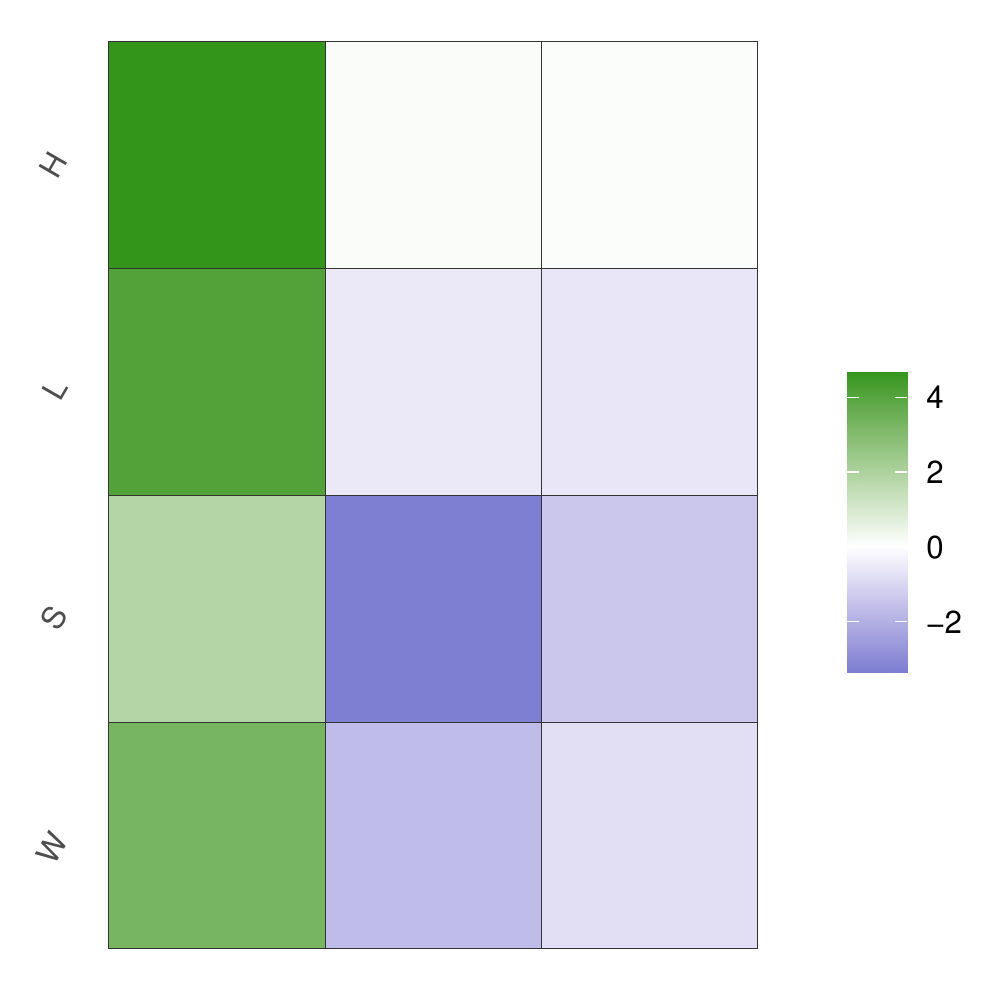}
\includegraphics[width = \textwidth]{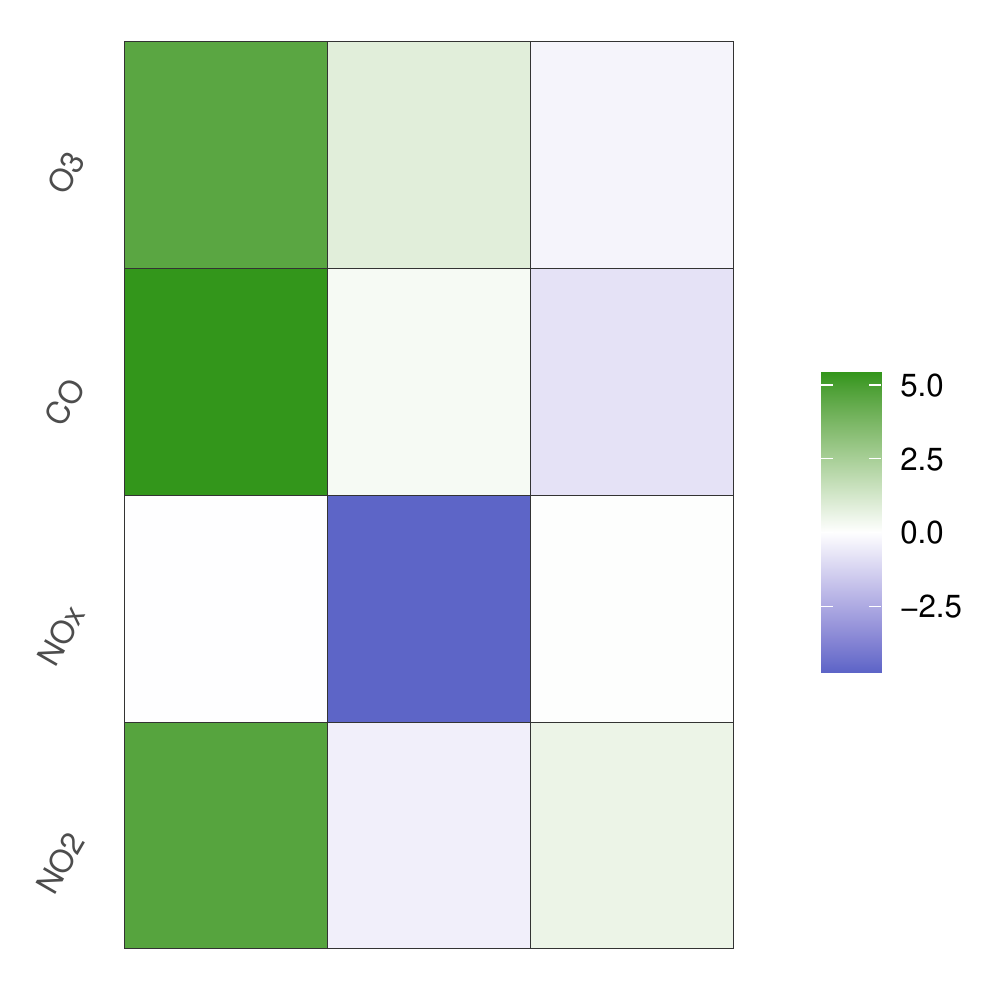}
\caption{Fix center \label{fig:loadings_fix}}
\end{subfigure}
\caption{Results associated with Ellipsoid-Gaussian. Left and right: the posterior mean of the post-processed factor loadings of $\Lambda$. Middle: the posterior mean (black) and the $95\%$ credible band (blue ribbon) of muscle mass (top) and the benzene level (bottom) as a function of shell mass and the CO level respectively, holding other covariates at their sample mean level. \label{fig:real_eg}}
\end{figure}

Figures~\ref{fig:horse_mussel_bfa} and the top row of ~\ref{fig:real_pairs} visualize the samples from the posterior predictive distributions associated with each model. Neither the copula factor model nor the Gaussian linear factor model captures the curvature at all. In fact, the copula model is only able to model the data mass in the center. Ellipsoid-Gaussian captures the curvature both when the center is fixed and when it is updated. While mixtures of factor analyzers also capture the curvature well, they use too many parameters---3 mixture components, with each containing 4 latent factors, which include 60 parameters for the factor loadings. This sacrifices the interpretability of the model. In contrast, our Ellipsoid-Gaussian model has a linear factor structure simplifying interpretation. To illustrate this, we visualize the posterior mean of the post-processed factor loadings in Fig.~\ref{fig:real_eg}. The first column encodes with the overall shell size (the sum of the length, the width, and the height) and the second column encodes the shell mass; this is the same regardless of whether the center is fixed or updated. 

Figure~\ref{fig:loadings_update} shows the relationship between the posterior mean of the muscle mass and the shell mass holding the other covariates at their sample mean level. There is a clear increasing relationship between shell mass and muscle mass. Relationship plots for other variables can be found in the Supplementary Material \citep{song23eg_sm}.

\subsection{Air quality data\label{sec:airquality}} 
Urban atmospheric pollutants are believed to have contributed to an increasing incidence of respiratory illnesses. To monitor urban air quality, a multi-sensor device was placed in a polluted area in an Italian city and recorded hourly averaged responses, reflecting concentrations of CO, total nitrogen oxides (NO$_x$), nitrogen dioxide (NO$_2$) and O$_3$ \citep{Vito2008OnFC}. The scatter plot matrix in Figure~\ref{fig:real_pairs_eg_fix} shows that the variables have curved dependence. The study tried to estimate the benzene concentration using the four measurements, with the benzene reference concentration provided by a conventional fixed station \citep{Vito2008OnFC}. The data set contains 8991 observations for each chemical. For Ellipsoid-Gaussian, we use $k = 3$ factors to model the data.

We were not able to run the package code for Gaussian copula factor model due to R session crashes. While we were able to complete a run of mixtures of factor analyzers for 10,000 iterations, it was rarely successful---the R session tended to crash after running for a long time. 
The completed run took around 305 minutes, whereas Ellipsoid-Gaussian took around 31 minutes (update center) and Gaussian linear factor model took 69 seconds to finish.

Figures~\ref{fig:real_pairs_eg_fix} and \ref{fig:real_pairs_eg_update} illustrate the posterior predictive distribution produced by the Ellipsoid-Gaussian when the center is fixed and updated respectively, with the fitted marginal distributions matching those of the original data and capturing the curvature. The posterior predictive distribution produced by Gaussian linear factor models does not match the curvature in the data while that by mixtures of factor analyzers contains outliers that are not present in the other models. Gaussian linear factor models underfit the data, while mixtures of factor analyzers produced an overly complex fit with twenty-one components, each with up to four latent factors. The Ellipsoid-Gaussian offers a good compromise between fit and interpretability.

Figure~\ref{fig:loadings_update} and \ref{fig:loadings_fix} visualize the posterior mean of the post-processed loadings matrix for when the center is updated and fixed respectively. Interestingly, both plots suggest the weighted average of O$_3$, CO and NO$_2$ concentration as one of the important latent factors. It also suggests NO$_x$ concentration itself as another important factor. Figure~\ref{fig:loadings_update} shows the relationship between the expected benzene concentration and the CO concentration holding all other variables at their sample mean level; there is a clear elbow shape, where the benzene level tends to remain stable when CO is at a lower level but increase linearly as CO continues to rise.


\section{Discussion}

In this article, we propose a simple and flexible von-Mises Fisher linear factor model to capture curved dependence in data, leading to a new class of Ellipsoid-Gaussian multivariate distributions. Simulating from the model results in points distributed about an ellipsoid, which is flexible enough to match curvature in many data sets. The use of a single factor loadings matrix facilitates dimension reduction and simple interpretation. In contrast, Gaussian linear factor models fail to characterize curvature and mixtures of factor analyzers 
are highly complex and can overfit the data, leading to spurious clusters and outliers in the posterior predictive.

In the absence of curved dependence in data, the model behaves like the routinely used Gaussian linear factor model, as we demonstrate both theoretically and empirically. We show how to marginalize out the latent factors and derive the density with respect to the Lebesgue measure. We also characterize various appealing properties of the distribution, such as that the marginal distribution of any sub-vector still follows an Ellipsoid-Gaussian. 

We devise a relatively efficient posterior sampler, despite the failing of the state-of-the-art algorithms, to sample from the associated posterior distribution. The sampler allows users to freely choose between updating and not updating the centers, with the code available in an R package.


There are a number of directions that would be interesting to pursue in future work. First, it is conceptually appealing to place shrinkage priors on the loadings, such as in \cite{kowal21orderspikeslab}. However, this leads to challenges in computation.  Currently, we sample the axes directions $U$ (i.e. the left singular vectors of $\Lambda$) and the axes lengths (i.e. the singular values of $\Lambda$). It is appealing to develop shrinkage priors for the axis directions to facilitate high-dimensional inferences, but the directions are constrained to the Stiefel manifold, making it unclear how to define appropriate priors.
Second, efficiency of fitting Ellipsoid-Gaussian models is constrained by computation for approximating the pseudo-normalising constant. A faster but still reliable method for approximating this term would be helpful. 

\begin{supplement}
\stitle{Supplementary Material: Curved factor analysis with the Ellipsoid-Gaussian distribution (DOI: XXXXXX/XXXXSUPP.pdf)}
\sdescription{provides additional results for the main manuscript. 
First, we provide proofs for Propositions \ref{prop:density}--\ref{prop:vmf_to_normal} and Lemma~\ref{lem:vmf_limit}. Then we discuss prior specification, gradient computation and sampling procedures for Ellipsoid-Gaussian. In the last part, we include additional results for the simulation study and applications.}
\end{supplement}
\bibliographystyle{ba}
\bibliography{main}

\begin{acks}[Acknowledgments]
This work was partially supported by the National Institute of Environmental Health Sciences of the United States National Institutes of Health under Grants R01ES027498 and R01ES028804.
\end{acks}

\end{document}